
%
\magnification = \magstep1
\baselineskip = 13pt
\font\BBrm=cmr12 scaled \magstep4

\font\Brm=cmr12 scaled \magstep2
\font\mrm=cmr10 scaled \magstep1
\font\BBit=cmti12 scaled \magstep4

\def\sk{\smallskip \noindent}
\def\msk{\medskip \noindent}
\def\bsk{\bigskip \noindent}

\font\bit= cmti12 scaled\magstephalf

\font\bsl= cmsl12

\font\bbd=cmbx12 scaled\magstep1

\font\bbrm=cmr12 scaled\magstephalf
\font\bbit=cmti12 scaled\magstephalf
\font\bb=msbm10
\font\refrm=cmr8
\font\refit=cmti8
\font\refsl=cmsl8
\font\refbf=cmbx8
\def\R{\hbox{\bb R}}
\def\G{\hbox{\bb G}}
\def\Z{\hbox{\bb Z}}
\def\N{\hbox{\bb N}}
\global\newcount\chcount
\global\newcount\subchcount
\global\newcount\glcount
\global\chcount=0
\def\glnum{{\global\advance\glcount by1}
{\number\chcount .\number\glcount}}
\def\eq{\eqno{(\glnum)}}
\def\ref#1!#2!#3!#4!#5!#6{\parindent=8mm\item{#1} \refrm #2\refsl #3\refit
#4\refbf\ #5\refrm\ #6}
\newif\iftitlepage \titlepagetrue
\newtoks\titlepagehead \titlepagehead={\hfil}
\newtoks\titlepagefoot \titlepagefoot={\hfil}
\newtoks\currentch \currentch={\hfil}
\newtoks\currentsubch \currentsubch={\hfil}
\newtoks\evenpagehead \newtoks\oddpagehead
\evenpagehead={\font\it=cmti8
               \vbox{\line{\bit\folio\hfil\font\bbrm=cmti8
                       \it\the\currentch\hfil}
                       \vskip-11pt\sk\line{\null\hrulefill}}}
\oddpagehead={\font\it=cmti8
              \vbox{\line{\hfil\it\the\currentsubch\hfil\bit\folio}
                       \vskip-11pt\sk\line{\null\hrulefill}}}
\newtoks\evenpagefoot \newtoks\oddpagefoot
\oddpagefoot={\hfil}
\headline={\iftitlepage\the\titlepagehead
\else\ifodd\pageno\the\oddpagehead
\else\the\evenpagehead\fi\fi}
\footline={\iftitlepage\the\titlepagefoot
\else\ifodd\pageno\the\oddpagefoot
\else\the\evenpagefoot\fi\fi}

\def\chhead#1{
\global\advance\chcount by1
\global\glcount=0
\null\bsk\centerline{\vbox{\line{\hfil{\bbd{#1}}\hfil}
                              \vskip-12pt
\line{\hfil\underbar{\phantom{\bbd{#1}}}\hfil}}}}
\def\subchhead#1{
\null\bigskip\line{{\indent\bsl #1}\hskip5pt\hrulefill}}
\centerline{\BBrm On {\BBit SW}-minimal models}
\bsk\sk
\centerline{\BBrm and {\BBit N=1} supersymmetric}
\bsk\sk
\centerline{\BBrm Quantum Toda-field theories}
\bsk\sk
\bsk\bsk\bsk\vfill
\centerline{\Brm Steffen Mallwitz\footnote{${}^\dagger$}{\rm
        \item{}Present address:
        \item{}Max-Planck-Institut f\"ur Physik komplexer
         Systeme, Bayreuther Str.40, 01187 Dresden}}
\bsk\bsk
\centerline{\mrm Abstract}
\bsk
Integrable $N$=1 supersymmetric Toda-field theories are determined by a
contragredient simple Super-Lie-Algebra (SSLA) with purely fermionic lowering
and raising operators. For the SSLA's $Osp(3|2)$ and $D(2|1;\alpha)$ we
construct explicitly the higher spin conserved currents and obtain free field
representations of the super $W$-algebras $SW(3/2,2)$ and $SW(3/2,3/2,2)$.
In constructing the corresponding series of minimal models
using covariant vertex operators, we find a necessary restriction
on the Cartan matrix of the SSLA, also for the general case.
Within this framework, this restriction claims that there be a
minimum of one non-vanishing element on the diagonal of the Cartan matrix.
This condition is without parallel in bosonic conformal field theory.
As a consequence only two series of SSLA's yield minimal
models, namely $Osp(2n|2n-1)$ and $Osp(2n|2n+1)$.
Subsequently some general aspects of degenerate representations of
$SW$-algebras, notably the fusion rules, are investigated.
As an application we discuss minimal models of $SW(3/2,2)$, which were
constructed with independent methods, in this framework.
Covariant formulation is used throughout this paper.
\vfill\vfill\bsk
\vbox{\line{Address: \hfil BONN-TH-94-04}
      \line{Physikalisches Institut, Nu{\ss}allee 12 \hfil hep-th/9405025}
      \line{53115 Bonn \hfil Bonn University}
      \line{Germany \hfil April 1994}
      \line{e-mail: steffen@avzw01.physik.uni-bonn.de \hfil}}
\filbreak
\eject
\vsize 10.0in
\titlepagefalse\pageno=1
\chhead{1. Introduction}
\bsk
The analysis of conformal quantum field theories in two dimensions [1]
by means of free field representations of the corresponding symmetry algebras
is considered to be a standard procedure today. It  gives insight into
fusion rules and degenerate representations as well as into minimal models.
In this way most known results concerning the Virasoro and Super-Virasoro
algebra can easily be reproduced [2-6].
Extensions of the Virasoro algebra by additional simple fields
are commonly called $W$-algebras, for a detailed review see [7].
For the case of the $A_n$-based $W_n$-algebras,
formulas for the conformal dimensions of the primary fields contained in
a rational model can be derived in essentially the same manner [8].
The generalization to arbitrary Casimir algebras can be found in [9].
In connection with classical Toda-field theories free field representations
emerge as a very natural tool and yield in conjunction with the Miura
transformation a means for the explicit construction of higher spin conserved
currents. In the case of a quantum theory the Miura transformation is not
at all obvious [10], though explicit calculations still enable one to
write down the conserved quantities.
\sk
In this paper this is performed for the case of two $N$=1 supersymmetric
Toda-field theories, namely for the ones based on $Osp(3|2)$ and
$D(2|1;\alpha)$. We find by a simple general calculation
strong evidence for a general rule indicating when one might expect not
to encounter any minimal models:
One simply has to check wether the Cartan matrix of the corresponding
simple Super-Lie-Algebra (SSLA) has non-vanishing diagonal elements. If there
is none, it is not possible to construct minimal models in the standard way.
As a consequence we find that only two series of SSLA's would be capable of
generating minimal models.
\msk
In the case of $Osp(3|2)$ the Cartan matrix has one non-vanishing diagonal
element and we demonstrate the applicability of our approach to all currently
known minimal models of $SW(3/2,2)$, which is the simplest
non-trivial $SW$-algebra and therefore compares to the
bosonic $W_3$-algebra. Subsequently we derive a
set of rules for determining the fields involved in a minimal model.
However, this set of rules is possibly not complete.
\sk
We would like to mention that many of the calculations that are presented
below, could only be performed with the help of a specially implemented
symbolic calculation package. This includes the computation of $N$=1 covariant
operator product expansions as well as the reduction of normal ordered
expressions to a standard form, confer also [11].\footnote{${}^1$}{This
symbolic calculation package OPESUSY for {$ Mathematica^{TM}$} is available
via anonymous FTP from the host avzw02.physik.uni-bonn.de.}
\sk
The organization of the paper is as follows:
After having set the stage in section two, we briefly review supersymmetric
Toda-field theories in section three, so that the results of the free field
constructions can be given in section four. In section five the basic
relations of $N$=1 representation theory are derived using the vertex operator
approach. These are then used to study the case of $Osp(3|2)$ or
$SW(3/2,2)$ in some detail.
\filbreak
\vsize7.5in
\chhead{2. Notations, a short account on superspace}
\bsk
This section serves to fix our notation. For details we refer the
reader to the general literature on the subject of supersymmetry as well
as to refs.~[4,~5] and references therein.
We work on superspace $S$, the extension of the two-dimensional
real vector space $\R^2$ with either euclidean or minkowskian metric
by a two-dimensional Grassmann space $\G^2$ spanned by two real Majorana
spinors $\theta_\alpha$.
Functions $\Phi(z,\theta)=\Phi(Z)$ defined on $S$ are called superfields.
Expanding in $\theta$ they take the following form
$$ \Phi(x,\theta) = \phi(x) + \bar \theta \psi(x) +
             {1 \over 2}\bar \theta \theta\,\xi(x)\ ,\eq $$
where $\bar\theta$ denotes the dirac conjugate of $\theta$.
Taking $S$ to be euclidean we define
$$ \theta = \theta_1 + \theta_2 ,\quad
\quad z=x_1 + x_2,   \quad Z=(z,\theta),
\quad D={\partial \over \partial\theta} + \theta{\partial \over \partial z},
\eq $$
as well as the corresponding bared quantities, e.g. $\bar\theta =
\theta_1 - \theta_2$
which bear a minus sign; and the super distance
$${Z_{ij}}^{n} =
\cases{(z_1-z_2-\theta_1 \theta_2)^n,&for $ n \in {\Z}$~;\cr
(\theta_1 - \theta_2)(z_1 -z_2 - \theta_1 \theta_2)^{(n-1/2)},
&for $ n \in {\Z}+{1\over 2}\ . $ \cr}  \eq
$$
Now we comment briefly on superconformal quantum field theories in two
dimensions [4,~5].
Their fields transform covariantly under the action of the infinitely
generated conformal group and respect in addition $N$=1 supersymmetry.
For a moment we consider chiral fields, i.e.~they depend only
on the unbared coordinates. The supersymmetric version of the energy
momentum tensor is a superfield $T = {1\over 2}G + \theta\,\tilde T$ where
the component fields are the ``ordinary'' bosonic energy momentum tensor
$\tilde T$ and the so-called supercurrent $G$.
The operator product expansion (OPE) of $T$ with itself gives the
Super-Virasoro algebra
$$ T(Z_1)\circ T(Z_2) = {c/6 \over {Z_{12}}^3} + {3/2\ T(Z_2)
\over{Z_{12}}^{3/2}} +
{1/2\ T'(Z_2) \over Z_{12}} + {T''(Z_2) \over{Z_{12}}^{1/2}}+ reg.\ . \eq$$
Fields $\Phi_i$ with well-defined superconformal dimension
are called primary fields, iff they satisfy
$$ T(Z_1)\circ \Phi_i(Z_2) = {\Delta_i \ \Phi_i(Z_2)\over {Z_{12}}^{3/2}} +
{1/2\ \Phi_i'(Z_2)\over Z_{12}} + {\Phi_i''(Z_2)\over {Z_{12}}^{1/2}}
+ reg.\ .\eq$$
The primary fields are a subclass of the quasiprimary fields, which
we do not define here [12]. Note that $T$ is a quasiprimary field.
A finite set of simple, (quasi)primary superfields including $T$ closing in
itself with the OPE as the binary associative operation is called a
$SW$-algebra. ``Closure'' means that the fields on the right hand
side of the OPE's may be the fields itself, their Super-derivatives
or normal ordered products thereof. Normal ordered products of superfields
are given by the normal ordering prescription for the component fields.
For more details on $SW$-algebras see e.g.~refs.~[7,~12,~13].
In section four we construct free field representations of certain
$SW$-algebras from non-trivial conserved currents of supersymmetric
Toda-field theories.
\vfill\eject
\chhead{3. {\bit N=1} Toda-field theories}
\bsk
In this section we introduce Toda-field theories and some of their basic
properties. For some relevant articles on this subject see
ref.~[15-25].
\sk
Let $K_{ij}$ be the Cartan matrix of a simple Super-Lie-Algebra $A$
with $r$~=~rank($A$) simple roots $\alpha_i$, fundamental weights $\mu_i$
and $\Phi$ a scalar superfield defined on $S$.
In the basis of the simple roots $\alpha_i$ $\Phi$ has $r$ component fields
$\Phi_i$.
The symmetric bilinear form on weight space is denoted by ''$\cdot$''.
Then the theory is defined on $S$ by the action
$$  {\cal S}[\Phi] =\int d\bar Z\,dZ\,
\Bigl({1\over 2} \bar D\Phi\cdot D\Phi +
{1\over \beta^2}\sum_{i\in {\cal F}}
\exp(\beta\,\alpha_i\cdot\Phi) +
{1\over \beta^2}\theta\,\bar\theta\,\sum_{i\in {\cal B}}
\exp(\beta\,\alpha_i\cdot\Phi)  \Bigr)\ ,
\eq $$
with $\beta$ being a coupling constant.
The sum over the bosonic roots {$\cal B$} spoils the supersymmetric
covariance, whence we restrict ourselves to SSLA's allowing for a system
of fermionic simple roots {$\cal F$} only.
We find a classification of these in refs.~[19,~28].
For completeness we list them here again: $sl(n+1|n)$,
$Osp(2n+1|2n)$, $Osp(2n-1|2n)$, $Osp(2n|2n-2)$, $Osp(2n|2n)$ and
$D(2|1;\alpha)$.
In preserving supersymmetry we also find that the action is now invariant
under superconformal transformations [19].
The equations of motion which follow from (3.1) are
$${\delta {\cal S} \over \delta \Phi_j}= 0
\qquad\Longleftrightarrow \qquad
\bar D D\Phi = \exp(\sum_i\beta\alpha_i\cdot\Phi)\ .
\eq $$
We will use them to construct the conserved higher
spin fields which are then shown to form a $SW$-algebra. In the rather
complex case of $D(2|1;\alpha)$ this is more straightforward than making a
general ansatz for a field with given spin-$n$ and subsequently fixing
the coefficients via equations obtained from the given OPE's [13] of the
$SW$-algebra.
\msk
In any decent quantum field theory the energy momentum tensor
is a conserved quantity.
As can be seen from a short computation, the quantity
$$\eqalign{ T =& -{1\over 2}\left(\sum_{i,j=1}^r K_{ij} N(D^2\Phi_i\,D\Phi_j) -
\sum_{i=1}^r ({1\over\beta} + \beta K_{ii}) D^3\Phi_i\right) \cr
=& -{1\over 2}N(D^2\Phi\cdot D\Phi) +
    {1\over\beta}\rho_q D^3\Phi\ , \cr
\rho_q =&\, {1\over 2}\,\sum_{i=1}^r (1 + \beta^2\,\alpha_i^2)\,\mu_i\ ,\cr}
\eq $$
is always conserved. ($\rho_q$ is the Weyl vector plus a
``quantum correction'', $N$ denotes the normal ordered product.)
This can be seen from the equations of motion eq.(3.2) and
the rules in appendix A.
Now we address the question of how to obtain the conserved fields
of higher spin, since that is what we are naturally interested in in order
to obtain a free field representation of the corresponding $SW$-algebra.
The existence of these fields is not trivial and related to
the integrability of the theory. The conformal spin content of the
conserved fields can be obtained from ref.~[22].
\sk
For explicit calculations we introduce the spin one-half field
$\Psi_i = D\,\Phi_i$, which serves as a ``basis'' for a spin-$n$ ansatz.
After a ``light-cone-like'' quantization [21] on a quantization surface
$\bar Z = const.$, the OPE
of a ``free'' field is
$$ \Psi_i(Z_1)\circ\Psi_j(Z_2) = -{K_{ij}^{-1}\over Z_{12}}+ reg.\ ,\eq$$
where the constant $\bar Z$-component has been omitted.
\sk
A conserved current of spin higher than that of the energy momentum tensor
is obtained by a general ansatz for a chiral spin-${n\over 2}$ field
of the following form
$$W = \sum_{{i_1,\dots,i_r,r\atop p(n)=(n_1,\dots,n_r)}}
a_{i_1,\dots,i_r,p(n)}\,N(D^{n_1}\Phi_{i_1},N(D^{n_2}\Phi_{i_2},
                            N( \dots,D^{n_r}\Phi_{i_r})))\ ,  \eq $$
where $p_r(n)$ denotes the monotonously ordered partitions of $n$ into $r$
integers. Note that eq.(3.5) is an expression in $\Psi_i$. The coefficients
are partially fixed by the condition that the application
of $\bar D$ annihilates it, i.e.~it is conserved under the
Toda dynamics. In this calculation one commutes the $\bar D$ with
the $D$'s until the equation of motion eq.(3.2) is applicable. The
derivatives of the exponentials are then evaluated and the
resulting terms, which are in general normal ordered products, are
rearranged to extract equations for the coefficients $\alpha_{p(n)}$.
This process was performed with the help of the aforementioned symbolic
calculation program. In the resulting expression no explicit $\bar Z$
dependence is retained.
\sk
By virtue of eq.(3.4) currents in terms of the $\Psi_i$
can be multiplied and may eventually give rise to a $SW$-algebra.
Classically eq.(3.4) is converted to a Poisson
bracket and the closure of the algebra of conserved currents to a classical
$SW$-algebra is a direct
consequence of the integrability of the Super Toda theory. This integrability
on the other hand stems from the underlying SSLA-structure, see also
refs.~[15,~19].
Of course in the quantum case too, integrability ensures the closure to a
quantum $SW$-algebra, but this is far less obvious and so far an unsolved
question for general SSLA's. Besides integrability we
are mainly interested in the free field representation of the $SW$-algebra in
order to investigate rational theories thereof.
\bsk
\filbreak

\chhead{4. Free field construction for {\bbrm {\bbit{SW}}(3/2,2)}
       and {\bbrm {\bbit SW}(3/2,3/2,2)}}
\bsk
Starting from the conserved currents of the Toda theory in
question we construct the simple fields of the corresponding
$SW$-algebra. The $SW(3/2,2)$ was first constructed in [26].
It has been, together with the $SW(3/2,3/2,2)$, presented using covariant
OPE's for the first time in [13] and we rely heavily on this kind of
representation. Finding a conserved
current, we expect it to be an additional simple, primary field of a
$SW$-algebra, though it is necessary to take its primary projection, first.
Since all equations we used so far are linear, the normalizations of the
fields are chosen in accordance with the non-linear OPE's of the $SW$-algebra
as given in [13].
\subchhead{Osp(3$|$2)}
\bsk
Our first example is $Osp(3|2)$-Toda-field theory.
The Cartan matrix is given by
$$
\pmatrix{1 & -1\cr -1& 0\cr}\ ,
\eq $$
so that the energy momentum tensor according to eq.(3.3) is
$$
T = -{1\over 2}\,\left(N(\Psi_1',\Psi_1) - N(\Psi_1',\Psi_2) -
N(\Psi_2',\Psi_1)\right) -
\left({1\over \beta}+ \beta\right) \Psi_1'' - {1\over \beta} \Psi_1''\ ,
\eq $$
where ``$\,'\,$'' stands for a covariant derivative $D$.
The associated Super-Virasoro algebra bears the central charge
$$ c = -(3 +{9 \over \beta^2})\ .\eq $$
Note that the sign of $c$ is quite formal at this point, since we
have not determined the interesting range of $\beta$, yet.
The most general ansatz for a chiral spin-2 field in two fields $\Psi_i$ is
$$
\tilde W=\sum_{i\leq j=1}^2 m_{ij}\,N(\Psi_i', \Psi_j') +
\sum_{ij=1}^2 n_{ij}\,N(\Psi_i'', \Psi_j) +
\sum_{{ijk=1\atop i<j}}^2 n_{ijk}  N(\Psi_i', N(\Psi_j,\Psi_k)) +
\sum_{i=1}^2 m_i \Psi_i'''\ ,
\eq $$
where for the moment we keep the $\bar Z$ dependance.
Requiring it to be conserved under the Toda dynamics,
i.e.~$\bar D\,\tilde W=0$, yields
$$
n_{22}= {{\cal N}\over 2 \beta},\quad n_{12}= ({1\over\beta} +
\beta)\ {\cal N},\quad
m_2=({1\over 2} + {1\over \beta^2})\ {\cal N},\quad n_{112}={-{\cal N}\over 2},
\quad n_{212}={{\cal N}\over 2}\ ,
\eq $$
with the normalization ${\cal N}$ still being a free parameter.
Since we want $\tilde W$ eventually to be the additional primary spin-2 field
of $SW(3/2,2)$, we introduce its primary projection
$$
 W= \tilde W + a\, D\, T\ ,
\eq $$
where $a$ evaluates to $a = -2\,{\cal N} (2+\beta^2)/(3 \beta)$.
With the choice of ${\cal N}$ as
$$
{\cal N}=\pm {3\ \beta\over \sqrt{5 \left( 2 - \beta \right)
\,\left( 2 + \beta \right) \,\left
( 1 + 2\,{\beta^2} \right)}}
\eq $$
the expressions for $T$ and $W$ are completely compatible
with the following OPE, which is cited from [13], so that we obtained a free
field representation of $SW(3/2,2)$.
$$\eqalign{W(Z_1)\circ W(Z_2) =&\quad{c\over {Z_{12}}^4}\ -
        \ {{6\over 5}\, T(Z_2)\over {Z_{12}}^{5/2}}\ +
    {C_1\ W(Z_2) - {2\over 5}\, T'(Z_2) \over {Z_{12}}^2}\ + \cr
        &{{1\over 2}C_1\, W'(Z_2) - {4\over 5}\, T''(Z_2) \over {Z_{12}}^{3/2}}
    \ + {{1\over 2}C_1\, W''(Z_2) - {1\over 5}\, T'''(Z_2) \over {Z_{12}}}\
+ \cr                &\hskip-85pt {C_2\, N(T',T)(Z_2) + C_3\, N(W,T)(Z_2) +
                 ({3\over 10}\, C_1 - {1\over 5}\, C_3) W'''(Z_2) -
            ({1\over 4}\, C_2 + {3\over 10}) T''''(Z_2) \over
          {Z_{12}}^{1/2}}\cr +
 ~reg.\ , }\eq $$
\sk
where the couplings are given by
$$
C_1^2 = {4 \over 5}{(5c+ 6)^2 \over (4c + 21)(c - 15)},\quad
C_2 = {108 \over 5}{1\over (4c+21)},\quad
C_3 = \pm {108 \over \sqrt{5}}{1\over \sqrt{(4c+21)(c-15)}}.\eq
$$
The free field representation for $SW(3/2,2)$ was first calculated in a
non-covariant component formulation in [14].
\sk
\subchhead{D(2$|$1;$\alpha$)}
\bsk
In this case three free fields are needed. The Cartan matrix which belongs
to a purely fermionic root system is
$$ \pmatrix{0& 1& -1 - \alpha\cr 1& 0& \alpha\cr -1-\alpha& \alpha& 0\cr}\ ,
\eq $$
so that the  energy momentum tensor takes the following form
$$ \eqalign{T~ =~& -{1\over 2}~\Bigl(N(\Psi_2',\Psi_1) +
(-1-\alpha)~N(\Psi_3',\Psi_1)) +
N(\Psi_1,\Psi_2')+\alpha~N(\Psi_3',\Psi_2) +\cr &(-1-\alpha)~N(\Psi_1',\Psi_3)+
\alpha~N(\Psi_2',\Psi_3)
 - {1\over\beta}~(\Psi_1''+\Psi_2''+\Psi_3'')\Bigr)\ . \cr}
\eq $$
Again we have a Super-Virasoro algebra with central charge
$$c = {3\over 2}\Bigl(3+{4\over \beta^2}(1+{1\over \alpha(1+\alpha)})\Bigr)\ .
\eq $$
For the spin-$3/2$ and spin-$2$ field we make the following most general ansatz
in three fields~$\Psi_i$
$$
\eqalign{
M~=~&\Bigl( \sum_{ij=1}^3 M_{ij}\ N(\Psi_i',\Psi_j) +
M_{123}\ N(\Psi_1,N(\Psi_2,\Psi_3))
+ \sum_{i=1}^3 M_i\  \Psi_i''\Bigr)\ , \cr
U~=~&\Bigl(\sum_{i\leq j=1}^3U_{ij}\ N(\Psi_i',\Psi_j')
   + \sum_{ij=1}^3 V_{ij}\ N(\Psi_i'',\Psi_j)
   + \sum_{i=1}^3 U_i\ \Psi_i'''   \cr
&+ \sum_{{ijk=1\atop j<k}}^3 U_{ijk}\ \ N(\Psi_i',N(\Psi_j,\Psi_k))\Bigr)\ .
\cr}
\eq $$
Evaluating $\bar D\,M=0$ as described in section three yields
$$
\eqalign{
   M_1 &= M_{31}/(\beta(1 + \alpha)),\cr
   M_{12} &= M_{32}/\alpha,\cr
   M_2 &= -M_{32}/(\beta\alpha),\cr
   M_{21} &= -M_{31}/(1 + \alpha),\cr}
\eqalign{
   M_{123} &= -\beta(M_{32} + M_{31}\alpha + M_{32}\alpha), \cr
   M_{13} &= -M_{31}\alpha - M_{32}(1 + \alpha)^2/\alpha,\cr
   M_{23} &= M_{32} + M_{32}\alpha + M_{31}\alpha^2/(1 + \alpha),\cr
   M_3 &= -M_{32}(1 + \alpha)/(\beta\alpha) -
           M_{31}\alpha/(\beta(1 +\alpha)),\cr}
  \hskip-10pt
  \eqalign{ M_{11} &= 0,\cr M_{22} &= 0,\cr M_{33} &= 0,\cr
		\phantom{M =0}\cr}
\eq $$
which contains two free parameters as expected, since $\bar D\,M=0$
is linear and $T$ is another solution independent of the primary field $M$.
The condition of primarity fixes $M_{31}$ to be
$$
 M_{31}=  {{ M_{32}\,{{\left( 1 +\alpha \right) }^2}\,\left( 2 +\alpha\right
) \,\left( 2 + {\beta^2}\,\alpha\right) }\over
    {\left( 1 -\alpha\right) \,{\alpha^2}\,\left( -2 +
{\beta^2}+{\beta^2}\,\alpha\right) }}\ .
\eq $$
For the spin-2-field the solutions are much more elaborate and since they are
not very elucidating here, they are presented in Appendix B.
Now, having obtained solutions for the conserved spin-3/2 and spin-2 fields
we investigate if their normalizations comply with
the nonlinear OPE's of ref.~[13], e.g.~the OPE of $M$ with itself
$$
\eqalign{ M(Z_1)\circ M(Z_2) =~&{c/6\over {Z_{12}}^3} +
             {{3\over 2}\, T(Z_2) + C_2\, M(Z_2)\over {Z_{12}}^{3/2}}
    + {{1\over 2}\, T'(Z_2) +
{1\over 3}\, C_2\, M'(Z_2) + C_3\, U(Z_2)\over {Z_{12}}}
\cr &+ {{1\over 2}\, C_3\, U'(Z_2) + T''(Z_2) +
{2\over 3}\, C_2\ M''(Z_2)\over {Z_{12}}^{1/2}} + ~reg.\ ,\cr  }
\eq $$
which results in
$$\eqalign{M_{32}^2 =~ & {{- {{\left( -1 +
\alpha \right) }^2}\,{\alpha^2}\,
        \left( -2 + {\beta^2} + {\beta^2}\,\alpha \right) }\over
    {6\,\left( 2 + {\beta^2} \right) \,{{\left( 1 + \alpha \right) }^2}\,
      \left( 2 + {\beta^2}\,\alpha \right) }} \ , \cr
C_2\,M_{32} =~& {{-2\,{{\left( -1 + \alpha \right) }^2}\,
      \left( 2 + 5\,\alpha + 2\,{\alpha^2} \right) }\over
    {3\,\left( 2 + {\beta^2} \right) \,{{\left( 1 + \alpha \right) }^2}\,
      \left( 2 + {\beta^2}\,\alpha \right) }} \ , \cr
C_3\,V_{23} =~&  {{-{{\left( 4 + 4\,\alpha + 3\,{\beta^2}\,\alpha +
4\,{\alpha^2} + 3\,{\beta^2}\,{\alpha^2} \right) }^2}}\over
    {18\,\left( 1 + \alpha \right) \,\left( 2 + {\beta^2}\,\alpha \right) \,
      \left( -2 + {\beta^2} + {\beta^2}\,\alpha \right) }}\ . \cr
}\eq $$
Analogously
$$\eqalign{M(Z_1)~\circ&~ U(Z_2) = {C_4\, M(Z_2)\over {Z_{12}}^2} +
          {C_5\, U(Z_2) + {1\over 3}\, C_4\, M'(Z_2)\over {Z_{12}}^{3/2}}
        + {{1\over 4}\, C_5\, U'(Z_2) +
    {1\over 3}\, C_4\, M''(Z_2)\over {Z_{12}}}\cr
             &+ {C_6\, (N(T,M)(Z_2)-{1\over 4}\, M(Z_2))+
         {1\over 2}\, C_5\, U''(Z_2) +
       {1\over 6}\, C_4\, M'''(Z_2)\over {Z_{12}}^{1/2}} + reg.\cr}\eq
$$
yields
$$\eqalign{M_{32}\,V_{23}=~&
     {{-\left( \left( 2 + {\beta^2} \right)\,\alpha\,M_{32}\,C_6 \right) }
   \over {6\,{\beta^2}}}\ ,
     \cr
   M_{32}\,C_4=~&
{{M_{32}\,C_6\,\left(4+4\,\alpha+3\,\beta^2\,\alpha+4\,\alpha^2+
 3\,{\beta^2}\,{\alpha^2} \right)}\over
 {6\,{\beta^2}\,\alpha\,\left( 1 + \alpha \right) }}\ ,
    \cr
   V_{23}\,C_5=~&
{{-4\,\left( 2 + {\beta^2} \right) \,M_{32}\,C_6\,
      \left( 2 + 5\,\alpha + 2\,{\alpha^2} \right) }\over
    {9\,{\beta^2}\,\alpha\,\left( -2 + {\beta^2} + {\beta^2}\,\alpha \right) }}
     \ ,\cr}\eq
$$
and last but not least
$$\eqalign{ U(Z_1)&\circ U(Z_2)= {-{c\over 10}\over {Z_{12}}^4} +
{C_7\,M(Z_2) - {6\over 5}\,T(Z_2)\over {Z_{12}}^{5/2}}  +
{C_8\,U(Z_2)+{1\over 3}\,C_7 M' - {2\over 5}\,T'(Z_2)\over {Z_{12}}^2} \cr &+
{C_8/2\ U'(Z_2) + 2/3\ C_7\ M''(Z_2) - 4/5\ T''(Z_2)\over {Z_{12}}^{3/2}}\cr&+
{C_8/2\ U''(Z_2) + 1/6\ C_7\ M'''(Z_2) - 1/5\ T'''(Z_2)\over {Z_{12}}}\cr&+
{C_9\,(2 N(T,T')(Z_2)-N(T',T)(Z_2)-1/4\,T''''(Z_2))\over {Z_{12}}^{1/2}}\cr&+
{C_{10}\ (N(M,M')(Z_2) ) + C_{11}\ N(T,M')(Z_2)  + C_{12}\ N(T,U)(Z_2))
\over {Z_{12}}^{1/2}}+ reg. \cr}\eq
$$
gives
$$\eqalign{
 V_{23}^2=~&
     {\alpha ^2\,\left( 2 + \beta ^2 \right) \,
          \left( 4 + 4\,\alpha  + 4\,\alpha ^2 +
            3\,\alpha \,\beta ^2 + 3\,\alpha ^2\,\beta ^2 \right)
          \over
        30\,\left( 4 - 2\,\beta ^2 - \alpha \,\beta ^4 -
            \alpha ^2\,\beta ^4 \right) } \ ,\cr
M_{32}\,C_7  =~&
     {4\,\left( -1 + \alpha  \right) ^2\,
          \left( 2 + 5\,\alpha  + 2\,\alpha ^2 \right) \over
        15\,\left( 1 + \alpha  \right) ^2\,
          \left( 2 + \beta ^2 \right) \,
          \left( 2 + \alpha \,\beta ^2 \right) }\ , \cr
V_{23}\,C_8 =~&
     {\left( 4 + 4\,\alpha  + 4\,\alpha ^2 + 3\,\alpha \,\beta ^2 +
            3\,\alpha ^2\,\beta ^2 \right) \,
          \left( 10 + 10\,\alpha  + 10\,\alpha ^2 +
            3\,\alpha \,\beta ^2 + 3\,\alpha ^2\,\beta ^2 \right)
          \over
        45\,\left( 1 + \alpha  \right) \,
          \left( -4 + 2\,\beta ^2 + \alpha \,\beta ^4 +
            \alpha ^2\,\beta ^4 \right) } \ .\cr}
\eq $$
The preceding calculations were again performed with the help of the
symbolic calculation program.
The coupling constants were first calculated in [13] and inserting
their values in eqs.(4.17), (4.19) and (4.21) gives a consistent solution
for $V_{23}$ and $M_{32}$. On the other hand the same set of equations
leads to exactly the coupling constants as in [13], so that
in some sense the structure of the $SW$-algebra, which we assumed
as known so far, is recovered in this fashion. We omitted the equations
which stem from the terms of order $Z_{12}^{-1/ 2}$ in eq.(4.20), since
they are too lengthy, but they do not change the picture.
In the next chapter it is shown that this free field representation for
$SW(3/2,3/2,2)$ does not support the construction of the corresponding series
of minimal models.
\bsk
\chhead{5. Representation Theory}
\subchhead{General considerations}
\bsk
The irreducible representations of the (Super-)Virasoro algebra are well-known,
and the extension of this to the $A_n$-series of bosonic $W$-algebras
has been achieved by Fateev and Lykyanov [8], simply by applying free field
techniques to the root systems of $A_n$. Much more elaborate are the
techniques used in [9], but the results cover all simple, classical Lie
algebras.
We will argue below that an application of the methods of ref.~[8]
using covariant vertex operators to the case of $N$=1 supersymmetry is
possible, although we encounter some qualitatively new features.
In the following we will restrict ourselves mainly to the Neveu-Schwarz-sector.
\sk
Let $W^{(k)}$ denote the simple\ (quasi)primary fields of the $SW(3/2,2)$-
and $SW(3/2,3/2,2)$-algebra,
where $k\in \{T,W\}$ or  $k\in \{T,M,U\}$, respectively. The modes of
the $W^{(k)}$ are given by
$$  W^{(k)} = \sum_{n\in\Z} W_{n\over 2}^{(k)}\, Z^{-n/2-\Delta_T(W^{(k)})}\ .
\eq $$
The power of $Z$ is obtained from eq.(2.3) by setting $Z_2$ equal to zero.
The Verma module ${\cal V}_{\Delta,c}$ is freely generated by the action of the
modes  $W_n^{(k)}$ with $n<0$ on $|\Delta \rangle$, a highest weight vector
satisfying
$$ W_{n\over 2}^{(k)}\,|\Delta\rangle = 0\ ,\quad \forall n > 0\quad
\hbox{and}\quad
W_0^{(k)}\,|\Delta\rangle = \Delta_k\,|\Delta\rangle\ . \eq$$
A state $|\xi_N\rangle$ is a null state in ${\cal V}_\Delta$
on level $N$, iff $|\xi_N\rangle$ is a highest weight vector and
$W^{(T)}_0\,|\xi_N\rangle = (\Delta_T + N)\,|\xi_N\rangle$.
Highest weight states are generated by applying primary fields at the origin
to the vacuum $|0\rangle$. The vertex operator
$V_\eta=\exp(\eta\cdot\Phi)$ is always a
primary field with superconformal dimension
$$ \Delta_T(V_\eta)=~-{1\over 2}\eta\cdot\eta +
{1\over \beta}\eta\cdot\rho_q\ =\Delta_T(\eta)\ . \eq$$
Here $\Phi$ is a free scalar superfield with conformal dimension zero and
$\eta$ is called the charge of the vertex operator.
We continue by introducing the notions of background charge and conjugation
of a charge. The background charge $\eta_0$ satisfies the relations
$$ \Delta_k(\eta)=\Delta_k(\eta_0-\eta),\qquad \forall k  \eq $$
and is related to $\rho_q$ by $\eta_0 = {2\over\beta}\rho_q$.
The conjugation is the involutive operation $\bar\eta = \eta_0 - \eta$
which preserves the superconformal dimension of $V_\eta$.
Next we introduce the so-called screening operators $Q_i$, which are
closely connected with the BRST-charge of refs.~[3,~6]. By definition
they commute with all simple (quasi-)primary fields $W^{(k)}$ of the
$SW$-algebra:
$[W^{(k)},Q_i]=0$. This induces $\Delta_T(Q_i)=0$, but the only two local
superfields with conformal dimension zero are equivalent to the identity.
Therefore we assume that the $Q_i$ are non-local and can be
represented in the following form
$$ Q_i = \oint_{\cal C} V_{\tilde\eta_i}\ , \eq$$
where the $\oint$ stands for ${1\over 2\pi i}\oint\, dz\, d\theta$.
Then the OPE of a field $W^{(k)}$ with $V_{\eta_i}$ is the total covariant
derivative with respect to $Z_2$ of some Laurent-polynomial in $Z_{12}$
$$ W^{(k)}(Z_1)\circ V_{\tilde\eta_i}(Z_2) = D_{Z_2} (\dots)\ . \eq$$
For $k = T$ this is equivalent to $V_{\tilde\eta_i}$ having superconformal
dimension one half.
Now, if $|\Delta(\eta)\rangle $ is a highest weight state, so is
$|\xi(\eta)\rangle = Q_i \,|\Delta(\eta)\rangle $, since
$$ W_{n\over 2}^{(k)}\,Q_i\,|\Delta(\eta)\rangle =
Q_i\,W_{n\over 2}^{(k)}\,|\Delta(\eta)\rangle = 0
\qquad \forall\, k,\ \forall\, n > 0\ .\eq$$
Since the computations involve the free field $\Phi$,
we evaluate this expression further (see also A.3),
$$\eqalign{|\xi(\eta)\rangle
=&\ Q_i\,|\Delta(\eta)\rangle\cr
=&\oint\ \exp(\tilde\eta_i\cdot\Phi(Z))
\circ \exp(\eta\cdot\Phi(0))\,|0\rangle \cr
=& \oint\ {Z\,}^{-\tilde\eta_i\cdot\eta}\,
\exp(\tilde\eta_i\cdot\Phi(Z)+\eta\cdot\Phi(0))\,|0\rangle\cr
=&{1\over [{m_i\over 2}]!}\,D_Z^{m_i}\Bigl(\exp(\tilde\eta_i\cdot\Phi(Z)+
\eta\cdot\Phi(0))\,|0\rangle\Bigr)\Bigr|_{Z=0}\ ,\cr}\eq$$
which is well-defined (and non-zero), iff
$$ \tilde\eta_i\cdot\eta =\,{m_i+1\over 2},\qquad m_i\in \N_0\ . \eq$$
$|\xi(\eta)\rangle$ is a highest weight state and contained
in a Verma module (the one of $|\Delta(\eta+\tilde\eta_i)$),
so it must be a null state.
For general states of the form
$$|\xi_{\tilde\eta_i\dots\tilde\eta_i}(\eta)\rangle =
\oint_{{\cal C}_m}\dots\oint_{{\cal C}_n} V_{\tilde\eta_i}\dots
V_{\tilde\eta_i}\,|\Delta(\eta_0-\bar\eta-l_i\,\tilde\eta_i)\rangle\ ,\eq $$
i.e.~$Q_i$ is applied $l_i$ times (with appropriately chosen
contours ${\cal C}_m$), the corresponding consistency
condition reads
$$\tilde\eta_i^2{l_i-1\over 2} + \tilde\eta_i\cdot
(\eta_0 -\bar\eta - l_i\,\tilde\eta_i) = {m_i+1\over 2}\ .\eq $$
We made use of $\exp(\eta_1\cdot\Phi(Z_1))\circ\exp(\eta_2\cdot\Phi(Z_2))=
Z_{12}^{-\eta_1\cdot\eta_2}\,\exp(\eta_1\cdot\Phi(Z_1)+\eta_2\cdot\Phi(Z_2))$,
where $\eta_1\cdot\eta_2$ takes integral values in the NS-sector,
so that $m_i$ is odd, while in the R-sector $m_i$ is even. Using eq.(5.3)
for $\tilde\eta_i$ we obtain
$$\tilde\eta_i\cdot\bar\eta =
{1 - m_i\over 2} +\tilde\eta_i^2{1-l_i\over 2}\ .\eq$$
\sk
Now we comment briefly on the spectrum of solutions of eqs.(5.6).
We rely on the (here unproven) fact that one part of the set of screening
charges is always given by
$\tilde\eta_i = \beta\,\alpha_i,\ 1\le i\le r$. All other screening charges
are assumed to be of the following form:
$\tilde\eta_i = \gamma_i\,\alpha_i,\ r+1\le i\le r+s$, where $\gamma_i$
will be in practice ${1\over\beta}$, see also [24,~25]. The ``pairing'' of
screening charges (linear dependency) will happen exactly when
$\alpha_i^2\not =0$, i.e.~the Cartan matrix of the SSLA has non-vanishing
elements on its diagonal.
\sk
Since the fundamental weights $\mu_i$ form a basis dual to the $\alpha_i$,
i.e.~$\alpha_i\cdot\mu_j=\delta_{ij},\ i,j\le r$ the final form of the allowed
charges of a vertex operator representing a highest weight of a degenerate
theory is
$$
\bar\eta= \sum_{i=1}^s\left(\beta{1-l_i\over 2}\right)\mu_i +
\sum_{i=1}^{r}\left({1\over \beta}{1-m_i\over 2}\right)\mu_i\ .
\eq $$
In generalization of a result of bosonic two-dimensional QFT, we
need the ``charge neutrality'' condition
$$ \sum_i \eta_i = \eta_0 \ ,\eq$$
for a $n$-point-function $\langle 0|V_{\eta_1}\cdots V_{\eta_n}|0\rangle$,
which is necessarily fulfilled, if it does not vanish identically.
This is equivalent to the restriction that it be independent of an ultraviolet
regularization.
By looking at three different ways of representing the three-point
function of primary fields by the corresponding (conjugated) vertex operators,
namely
$$\langle 0|V_{\bar\gamma}\, V_\delta\, V_\epsilon\, Q_{\tilde\eta_i}
\dots Q_{\tilde\eta_j}|0\rangle\quad
\langle 0|V_{\gamma}\, V_{\bar\delta}\, V_\epsilon\, Q_{\tilde\eta_i}
\dots Q_{\tilde\eta_j}|0\rangle\quad
\langle 0|V_{\gamma}\, V_\delta\, V_{\bar\epsilon}\, Q_{\tilde\eta_i}
\dots Q_{\tilde\eta_j}|0\rangle ,\eq$$
one obtains with eq.(5.14) equations of the following form:
$$ \eta_0-\gamma+\delta+\epsilon+\sum_i n_i\, \beta\alpha_i
   + \sum_i \hat n_i\,{1\over \beta} \alpha_i = \eta_0\ ,\eq$$
where $n_i$ and $\hat n_i$ denote the respective number of screening charges.
Now $\gamma$, $\delta$, $\epsilon$ are, according to
eq.(5.13), charges belonging to degenerate theories, e.g.
$$ \gamma = \sum_{i=1}^s\beta\,{\hat l_{ci}}\,\alpha_i +
\sum_{i=1}^{r}{1\over \beta}\,{\hat m_{ci}}\,\alpha_i\ ,
\eq $$
where we defined $\hat l_{cj}=\left({1-l_{ci}\over 2}\right)\,K_{ij}^{-1}$
and made use of $\mu_i = K_{ij}^{-1}\alpha_j$.
We introduced the additional subscript $c$ to denote that the coefficient
belongs to the charge $\gamma$ and the same was done for the $m_i$'s and also
for the charges $\delta$ and $\epsilon$ where the subscripts are $d$, $e$,
respectively.
This leads to equations in the coefficients $l_{i}$ and $m_{i}$, namely
$$  \beta^2{\hat l_{ci}}  + \hat m_{ci} = \,
             \beta^2\left(\hat l_{di}+\hat l_{ei}\right) +
            {\hat m_{di}+\hat m_{ei}} +
                 n_i\,\beta^2 + \hat n_i\ .\eq$$
We solve a notational problem by
setting $l_i = 1$ for $s+1\le i\le r+s$, since in general $s< r$.
Of course, this procedure is to be repeated for the other three-point
functions of (5.15), so that if we introduce the abbreviation
$k_i = \beta^2\,\hat l_i + \hat m_i$, we find a set of equations
$$
\eqalign{ { k_{ci}} & =
             \left( k_{di}+ k_{ei}\right) + n_i\,\beta^2 + \hat n_i\ , \cr
            { k_{ci}}  & =
            \left({ k_{di}- k_{ei}} \right) - n_i\,\beta^2 - \hat n_i\ ,\cr
	{ k_{ci}}  & =
            \left({ k_{ei}- k_{di}} \right) - n_i\,\beta^2 - \hat n_i\ .\cr}
\eq $$From
the above remarks it is clear that $\hat n_i$ can be identically zero
for some $i$, so that together with strictly positive $n_i$, the above set
of equations restricts the possible values of $k_{ci}$. These restrictions
form the desired fusion rules, which are a first step towards the exact
fusion.
At this point it is interesting to note that this set of fusion rules does
not only deviate from the standard way of writing them [4], but they are
also inequivalent for $SW$-algebras. For the Super-Virasoro-algebra
our version is compatible with the standard one. The difference between the
two is, that we do not impose relations by equating the coefficients
of powers of $\beta$.
In deriving minimal models
within the free field framework, one takes advantage of the periodicity
of the conformal dimensions $\Delta_k$ in the parameters $l_i$ and $m_i$.
This can be expressed by a relation of the following kind
$$ p_i\,\tilde\eta_i + q_i\,\tilde\eta_{r+i} = 0\ . \eq $$
Of course we may run into the problem that $\tilde\eta_{r+i}$ does not
exist for some $i$, that is we face a new situation, unknown in bosonic CFT.
The possible range of values for $l_i$ and $m_i$ is usually
determined by these numbers $p_i$ and $q_i$, so that the question arises
how this is accomplished for the case of a ``non-paired'' screening charge
$\tilde\eta_i$. We do not answer this question in this general context,
but refer to the next section, where we derive some rules inductively.
\sk
Another approach to perform a free field construction is to impose that the
$SW$-fields commute with the operators $\exp(\beta\alpha_i\,\Phi)$
$$[W^{(k)},\exp(\beta\alpha_i\,\Phi)] = 0 \ .\eq$$
In this way it is explicit that the charges $\beta\alpha_i$ give rise
to screening operators.
In our approach we also obtain these solutions, but we will not
always get their ``partners'' $\exp({1\over \beta}\alpha_i\,\Phi)$
as it is in the bosonic case, confer ref.~[24,~25].
\filbreak
\subchhead{SW(3/2,~2)}
\bsk
Partial results of this example were presented in ref.~[18], but since the OPE
given there differs from eq.(4.8), we performed an independent calculation.
Nevertheless we obtain the same results for the screening charges as
in ref.~[18].
The superconformal dimension of a vertex operator $V_\eta$ for this example is
$$\eqalign{\Delta_T(V_\eta)=&-{1\over 2}
\eta\cdot\eta + {1\over \beta}\eta\cdot\rho_q \cr
=&{-{{\eta_1^2}}\over 2}+\eta_1\,\eta_2+{\eta_1+\eta_2\over{2\,\beta}}
+{{\eta_1\,\beta}\over 2}\ , \cr
\hbox{with}\quad \rho_q =&-{1\over 2}\left(1,\beta^2 +  2 \right)\ .\cr} \eq$$
The $W$-dimension is
$$\eqalign{\Delta_W(V_\eta)=
{\cal N}\,&\biggl({{3\,{\eta_2^2}\,\beta+\eta_2\,(2+{\beta^2} )  +
      {\eta_1^2}\,( 4\,\beta + 2\,{\beta^3} )\over{6\,{\beta^2}}}}\  +\cr
     & {{\eta_1\,( -4 - 6\,{\beta^2} - 2\,{\beta^4} +
 \eta_2\,(-8\,\beta-4\,{\beta^3}))}\over{6\,{\beta^2}}}\biggr)\ .\cr
}\eq$$
The background charge $\eta_0$ satisfies $\Delta_T(V_{\eta_0})=0$ and
$\Delta_W(V_{\eta_0})= 0$, which determines it to be
$$\eta_0=\left(-{1\over \beta},-\left({2\over \beta}+\beta\right)\right)
     = {2\over \beta}\rho_q\ . \eq $$
It also satisfies $\Delta_T(V_{\eta_0-\eta})=\Delta_T(V_{\eta})$ and
$ \Delta_W(V_{\eta_0-\eta})=\Delta_W(V_{\eta}) $ as was to be expected.
For the screening charges $\eta_i$ we evaluate
$$ W(Z_1)\circ V_{\tilde\eta_i}(Z_2)=\,D_{Z_2}(\dots)+reg.\ ,\eq$$
and obtain three different solutions
$$\tilde\eta_1=\beta\,\alpha_1\ ,\quad
\tilde\eta_2 = \beta\,\alpha_2\ ,\quad
\tilde\eta_3 ={1\over\beta}\,\alpha_1\ ,\eq $$
which yield the correct superconformal dimension $\Delta_T(V_{\tilde\eta_i})
=\, {1\over 2}$.
This is quite surprising, since we expected four solutions at this stage.
One immediate consequence of this is that eq.(5.20) involves only two of the
three screening charges
$$ p\,\tilde\eta_1 + q\,\tilde\eta_3 = 0\qquad
\Longleftrightarrow \qquad {1\over \beta^2} = -{p\over q}\ ,\eq$$
which then gives rise to the following formula for the central charge of
minimal models
$$ c = -3\,(1 - 3\,{p\over q})\ .\eq$$
Note the linear appearance of ${p\over q}$ and since $c\le 3$ is necesssary for
minimal models [27] we find
${p\over q}\le{2\over 3}$. From eqs.~(5.17), (5.26) and (5.27) we find the
parametrisation of superconformal dimensions
$$\eqalign{\Delta_T(l_1,m_1,m_2) &=
{{-2\,q + 2\,m_2\,l_1\,q + 3\,p - {m_2^2}\,p - 2\,m_2\,m_1\,p}\over {8\,q}}\cr
&= {c(p,q)-3\over 24}+ {n\,(r\,q-s\,p)\over 8\,q}\ . \cr
&\ n= m_2,\ r=2\,l_1,\ s=2\,m_1 + m_2\cr} \eq $$From
this we observe $\Delta_T(l_1+p,m_1+q,m_2) = \Delta_T(l_1,m_1,m_2) $,
which actually reflects a symmetry of the charges of a minimal model as
is explained by eq.(5.20). It can be combined with the symmetry of
the weights $\Delta_k(\eta)$ as in eq.(5.4), if we formally introduce the
parameter $l_2$ corresponding to the (unphysical) screening charge
${1\over \beta}\eta_2$. Then the above shift operation extends also to the
pair $(l_2,m_2)$. By setting
$\alpha = \beta\,\lambda + {1\over\beta}\,\lambda' $, the transformation
$$
\eqalign{  	\lambda &\rightarrow p\,\rho_q - \lambda\ , \cr
		\lambda' &\rightarrow (q+2)\rho_q - \lambda'\ , \cr}
\eq $$
implies
$$
\eqalign{\alpha\ \rightarrow\qquad &\beta\,p\,\rho_q -\beta\,\lambda
         +{1\over \beta}\,(q+2)\,\rho_q -{1\over \beta}\,\lambda' \cr
&=(\beta\,p+{1\over \beta}\,q)\,\rho_q - \alpha +{2\over \beta}\,\rho_q \cr
&= \eta_0 - \alpha\ , \cr}
\eq $$
which is equivalent to the index shifts
$$
\eqalign{l_1\rightarrow (q-p-l_1+2)\ ,\qquad& l_2 \rightarrow (2-p-l_2)\ ,\cr
m_1\rightarrow\left({(2+q)q\over p}-q-m_1\right)
\ ,\qquad& m_2 \rightarrow (-q-m_2)\ . \cr}
\eq $$
Of course $l_2=1$ was implicitly assumed so far and ${(2+q)\over p}$ must
be an integral number which yields $p\ne 2$ if $p$ and $q$ are relatively
prime integers.
\bsk
So far there are four known $SW(3/2,2)$ minimal
models [29,~30]]:
\bsk
\vbox{
\hbox{
\vrule \hskip 1pt
\vbox{ \offinterlineskip
\def\tablespace{height2pt&\omit&&\omit&&\omit&&\omit&&\omit&\cr}
\def\tablerule{ \tablespace
                \noalign{\hrule}
                \tablespace        }
\hrule
\halign{&\vrule#&
  \strut\hskip 4pt\hfil#\hfil\hskip 4pt\cr
\tablespace
& $\left(\Delta_T,\, \Delta_W^2\right)$ &&
$\left(0         ,          \,0\right)$ &&
$\left({1\over 5},-{1\over 25}\right)$ &&
$\left({1\over 5}, {1\over 25}\right)$ &&
$\left(-{1\over 10}, 0\right)$  &\cr
\tablespace
\tablerule
& $\left(l_1,\,m_1,\,m_2\right)$ &&
  $\left(1,1,1\right)$ &&
  $\left(2,2,1\right)$ &&
  $\left(1,1,3\right)$ &&
  $\left(1,3,1\right)$ & \cr
\tablespace\tablespace
}
\hrule}\hskip 1pt \vrule}
\hbox{\ Table 1: $ (p,q)=(1,5)\qquad c=-{6 \over 5} $}
}
\bsk
\overfullrule3pt
\vbox{
\hbox{
\vrule \hskip 1pt
\vbox{ \offinterlineskip
\def\tablespace{height2pt&\omit&&\omit&&\omit&&\omit&&\omit&&\omit&&\omit&\cr}
\def\tablerule{ \tablespace
                \noalign{\hrule}
                \tablespace        }
\hrule
\halign{&\vrule#&
\strut\hskip 4pt\hfil#\hfil\hskip 4pt\cr
\tablespace
& $\left(\Delta_T,\, \Delta_W\right)$ &&
$\left(0,\,0\right)$ &&
$\left(0,\,{3\over 22}\right)$ &&
$\left({1\over 12},-{1\over 24}\right)$ &&
$\left(-{1\over 6}, {1\over 66}\right)$ &&
$\left({3\over 4},{39\over 88}\right)$ &&
$\left(-{1\over 4}, -{1\over 88}\right)$
 &\cr
\tablespace
\tablerule
& $ \left(l_1,\,m_1,\,m_2\right)$ &&
$\left(1,1,1\right)$ &&
$\left(0,1,3\right)$ &&
$\left(1,3,1\right)$ &&
$\left(0,3,1\right)$ &&
$\left(1,1,3\right)$ &&
$\left(0,1,1\right)$
 & \cr
\tablespace\tablespace
}
\hrule}\hskip 1pt \vrule}
\hbox{\ Table 2: $ (p,q)=(-1,6)\qquad c=-{9 \over 2} $}
}
\footnote{${}^a$}{In tables 2 - 4 we divided $\Delta_W$ by $C_1$}
\bsk
\overfullrule3pt
\vbox{
\hbox{
\vrule \hskip 1pt
\vbox{ \offinterlineskip
\def\tablespace{height2pt&\omit&&\omit&&\omit&&\omit&&\omit&&\omit&\cr}
\def\tablerule{ \tablespace
                \noalign{\hrule}
                \tablespace        }
\def\tablesmallspace{height1pt&\omit&&\omit&&\omit&&\omit&&\omit&&\omit&\cr}
\def\tableinterrule{ \tablespace
                \noalign{\hrule}
		\tablesmallspace
		\noalign{\hrule}
                \tablespace        }
\hrule
\halign{&\vrule#&
\strut\hskip 4pt\hfil#\hfil\hskip 4pt\cr
\tablespace
& $\left(\Delta_T,\, \Delta_W\right)$ &&
$\left(0,\,0\right)$ &&
$\left({1\over 14},{1\over 2}\right)$ &&
$\left(-{1\over 14}, {1\over 28}\right)$ &&
$\left({13\over 14},{11\over 7}\right)$ &&
$\left({3\over 7}, {39\over 28}\right)$
&\cr
\tablespace
\tablerule
& $ \left(l_1,\,m_1,\,m_2\right)$ &&
$\left(1,1,1\right)$ &&
$\left(1,3,3\right)$ &&
$\left(1,3,1\right)$ &&
$\left(2,2,3\right)$ &&
$\left(1,1,5\right)$
& \cr
\tablespace
\tableinterrule
& $\left(\Delta_T,\, \Delta_W\right)$ &&
$\left({2\over 7},{13\over 14}\right)$ &&
$\left({3\over 14}, {-3\over 4}\right)$ &&
$\left({1\over 7}, -{11\over 28}\right)$ &&
$\left(-{1\over 7}, -{1\over 28}\right)$ &&
$ $ &
\cr
\tablespace
\tablerule
& $ \left(l_1,\,m_1,\,m_2\right)$ &&
$\left(1,1,3\right)$ &&
$\left(2,2,1\right)$ &&
$\left(1,5,1\right)$ &&
$\left(2,4,1\right)$ &&
$ $
& \cr
\tablespace\tablespace
}
\hrule}\hskip 1pt \vrule}
\hbox{\ Table 3: $ (p,q)=(1,7)\qquad c=-{12 \over 7} $}
}
\bsk
\overfullrule3pt
\vbox{
\hbox{
\vrule \hskip 1pt
\vbox{ \offinterlineskip
\def\tablespace{height2pt&\omit&&\omit&&\omit&&\omit&&\omit&&\omit&&\omit&\cr}
\def\tablerule{ \tablespace
                \noalign{\hrule}
                \tablespace        }
\def\tablesmallspace{
height1pt&\omit&&\omit&&\omit&&\omit&&\omit&&\omit&&\omit&\cr}
\def\tableinterrule{ \tablespace
                \noalign{\hrule}
		\tablesmallspace
		\noalign{\hrule}
                \tablespace        }
\hrule
\halign{&\vrule#&
\strut\hskip 4pt\hfil#\hfil\hskip 4pt\cr
\tablespace
& $\left(\Delta_T,\, \Delta_W\right)$ &&
$\left(0,\,0\right)$ &&
$\left({7\over 8},{1\over 2}\right)$ &&
$\left({3\over 2}, {51\over 52}\right)$ &&
$\left(-{1\over 16},{43\over 416}\right)$ &&
$\left({1\over 16},-{1\over 32} \right)$ &&
$\left(-{1\over 4},-{1\over 52}\right)$ &
\cr
\tablespace
\tablerule
& $ \left(l_1,\,m_1,\,m_2\right)$ &&
$\left(1,1,1\right)$ &&
$\left(1,3,3\right)$ &&
$\left(1,1,5\right)$ &&
$ \left(0,1,3\right)$ &&
$\left(1,3,1\right)$ &&
$\left(0,1,1\right)$ &
\cr
\tablespace
\tableinterrule
& $\left(\Delta_T,\, \Delta_W\right)$ &&
$\left(-{1\over 8},{1\over 52}\right)$ &&
$\left({1\over 4},{4\over 13} \right)$ &&
$\left({11\over 16},{187\over 416}\right)$ &&
$\left(-{3\over 16},{3\over 416} \right)$ &&
$\left({1\over 8},{11\over 52}\right)$ &&
$\left({1\over 8},-{1\over 13} \right)$ &
\cr
\tablespace
\tablerule
& $ \left(l_1,\,m_1,\,m_2\right)$ &&
$\left(0,5,1\right)$ &&
$\left(0,1,5\right)$ &&
$ \left(1,1,3\right)$ &&
$\left(0,3,1\right)$ &&
$\left(0,3,3\right)$ &&
$\left(1,5,1\right)$ &
\cr
\tablespace\tablespace
}
\hrule}\hskip 1pt \vrule}
\hbox{\ Table 4: $ (p,q)=(-1,8)\qquad c=-{33 \over 8} $}
}
\bsk
Of course it is arbitrary up to now, whether we place the minus sign on $q$ or
$p$. Since the parameters $l_1$ and $m_1$ are periodic in $(p,q)$, we choose
the smallest non-negative value for them.
As can be seen from the above examples, we have to distinguish two
cases, namely $q$ odd or even.
Let us empirically deduce some ``selection rules''.
In the case $q$ odd the sum of $l_1$ and $m_1$ is always even, which is
also true for $l_2 + m_2$, if $l_2$ in accordance with the previous remarks
is set equal to one. This can be understood in context with eq.(5.11).
Furthermore we have $\sum_i m_i+l_i \le p+q$ and $1 \le l_i \le m_i$
(which is void for $i=2$) and after closer inspection one finds that this
already determines all possible fields, except for $(p,q) = (1,7)$. Here
this rule predicts a field with parametrization $(3,3,1)$ and dimensions
$\left({3\over 7}, \pm{69\over 28}\right)$, which is not observed.
There are two possibilities to interpret this. Either the above rules
are not complete yet, i.e.~they should be complemented by a rule stating
for example that $l_1$ and $m_1$ cannot be the same odd number other than
one, or we encounter a special case where the $(3,3,1)$-field belongs in
principle to the multiplet, but drops out of the fusion, because the
respective fusion coefficients are zero. The latter could not be clarified
within the present approach. Nevertheless, for $p$ positive the selection rules
are very similar to the ones of the $N=1$ Super-Virasoro algebra.
\sk
For the case $q$ even, since $(l_1,m_1)$ is defined only modulo $(p,q)$,
$l_1 + m_1$ is not necessarily even any more, though $l_2 + m_2$ is.
Also we find a pairing in the sense that if there is a field
of the form $(1,m_1,m_2)$ there is one of the form $(0,m_1,m_2)$ and
vice versa, which predicts an even number of fields. This seems to be the
only allowed value of $l_1$, while the range of $m_1$ and $m_2$ is
restricted by $1 \le m_1 + m_2 \le p+q $ with $m_1$ and $m_2$ odd.
Again, this is sufficient to parameterize all fields in the above
example, but it is of course no proof for higher $q$. After all it is not
clear if we chose the optimal parametrization of the conformal dimensions,
though our choice is motivated by the construction of the null fields.
\sk From
the above minimal models one can also empirically deduce the effective
central charge:
$c_{eff}=c-24\,\Delta_{min}$, namely: $ c_{eff}= 3-{9\over q}$, for
$ c=c\,(\pm 1,q)$. This is equivalent to
$$\Delta_{min}=\cases{{1\over 12}\,c,&for $c=c(1,q)$,\quad $q$ odd;\cr
 -{1\over 4},&for $c=c(-1,q) $, $q$ even. \cr}  \eq
$$
The next minimal model would be expected at $c=-2$ with $(p,q)=(1,9)$.
Unfortunately the explicit construction of this model is today beyond the
computational possibilities.
Apart from the above relations an explanation of the fact why $|p|=1$
and why the sign changes with $q$ being even or odd is still due.
\sk
It is not difficult to generalize eq.(5.32) to the Ramond-sector,
since the two spin fields which have to be introduced then contribute twice
the conformal dimension of ${1\over 16}$.
Indeed we can parametrize the $\Delta_T$-values of the above models for the
R-sector [29,~30], but since we do not know the OPE of a spin-field with
the free field $\Phi$ we cannot compute their respective $W$-dimensions.
We give the parametrizations of the two minimal models for which the
Ramond-sector is known.
\bsk
\vbox{
\hbox{
\vrule \hskip 1pt
\vbox{ \offinterlineskip
\def\tablespace{height2pt&\omit&&\omit&&\omit&&\omit&\cr}
\def\tablerule{ \tablespace
                \noalign{\hrule}
                \tablespace        }
\hrule
\halign{&\vrule#&
  \strut\hskip 4pt\hfil#\hfil\hskip 4pt\cr
\tablespace
& $\Delta_T$ &&
${3\over 4}$ &&
${3\over 20}$ &&
$-{1\over 20}$ &\cr
\tablespace
\tablerule
& $\left(l_1,\,m_1,\,m_2\right)$ &&
  $\left(2,1,2\right)$ &&
  $\left(1,2,2\right)$ &&
  $\left(x,y,0\right)$ & \cr
\tablespace\tablespace
}
\hrule}\hskip 1pt \vrule}
\hbox{\ Table 5: $ (p,q)=(1,5)\qquad c=-{6 \over 5} $
}
}
\bsk
\overfullrule3pt
\vbox{
\hbox{
\vrule \hskip 1pt
\vbox{ \offinterlineskip
\def\tablespace{height2pt&\omit&&\omit&&\omit&&\omit&&\omit&&\omit&&\omit&\cr}
\def\tablerule{ \tablespace
                \noalign{\hrule}
                \tablespace        }
\hrule
\halign{&\vrule#&
\strut\hskip 4pt\hfil#\hfil\hskip 4pt\cr
\tablespace
&$\Delta_T$ &&
${7\over 48}$ &&
${59\over 48}$ &&
$-{5\over 48}$ &&
${1\over 16}$ &&
${9\over 16}$ &&
$-{3\over 16}$
 &\cr
\tablespace
\tablerule
& $ \left(l_1,\,m_1,\,m_2\right)$ &&
$\left(0,0,4\right)$ &&
$\left(2,4,2\right)$ &&
$\left(0,0,2\right)$ &&
$\left(0,2,2\right)$ &&
$\left(0,0,6\right)$ &&
$\left(x,y,0\right)$
 & \cr
\tablespace\tablespace
}
\hrule}\hskip 1pt \vrule}
\hbox{\ Table 6: $ (p,q)=(-1,6)\qquad c=-{9 \over 2} $}
}
\bsk
\overfullrule3pt
For $q$ odd the selection rule $l_i + m_i$ odd and $\sum_im_i+l_i\le p+q$
seems to work. This would be complementary to the NS-sector, except that
$1\le l_i \le m_i $ is not satisfied. This leads in general to more fields
in the R-sector.
\sk
For $q$ even we find even values for $l_i$ and $m_i$, but the selection
rules are not very clear. Common to the two models is that the field
of dimension $\Delta_T = c/24$ is parameterized by $(x,y,0)$, where $x$ and
$y$ are arbitrary, as long as $m_2$ is equal to zero.
\bsk
Now, let us have a closer look at the fusion of the minimal models at
$c = -6/5$ and $c = -9/2$.
The model at $c$ = -6/5 is a Super Virasoro minimal model and therefore
its fusion is known exactly [31]. The non-zero fusion coefficients are all
equal to one.
\bsk
\vbox{
\hbox{
\vrule \hskip 1pt
\vbox{ \offinterlineskip
\def\tablespace{height2pt&\omit&&\omit&&\omit&&\omit&&\omit&\cr}
\def\tablerule{ \tablespace
                \noalign{\hrule}
                \tablespace        }
\def\tablesmallspace{height1pt&\omit&&\omit&&\omit&&\omit&&\omit&\cr}
\def\tableinterrule{ \tablespace
                \noalign{\hrule}
		\tablesmallspace
		\noalign{\hrule}
                \tablespace        }
\hrule
\halign{&\vrule#&
\strut\hskip 4pt\hfil#\hfil\hskip 4pt\cr
\tablespace
& $\left(\Delta_T,\, \Delta_W^2\right)$  &&
$\left(0,\,0\right)$ &&
$\left({1\over 5},-{1\over 25}\right)$ &&
$\left({1\over 5}, {1\over 25}\right)$ &&
$\left(-{1\over 10}, 0 \right)$
&\cr
\tablespace
\tableinterrule
& $\left(0,\,0\right)$ &&
$\Phi_1$ &&
$\Phi_2$ &&
$\Phi_3$ &&
$\Phi_4$
& \cr
\tablespace
\tablerule
& $\left({1\over 5},-{1\over 25}\right)$ &&
$\Phi_2 $ &&
$\Phi_1,\Phi_2 $  &&
$\Phi_4 $  &&
$\Phi_3,\Phi_4 $ &
\cr
\tablespace
\tablerule
&$ \left({1\over 5}, {1\over 25}\right)$ &&
$\Phi_3$ &&
$\Phi_4$ &&
$\Phi_1,\Phi_3$ &&
$\Phi_2,\Phi_4$ & \cr
\tablespace
\tablerule
& $\left(-{1\over 10}, 0 \right)$  &&
$\Phi_4 $ &&
$\Phi_3,\Phi_4 $ &&
$\Phi_2,\Phi_4 $ &&
$\Phi_1,\Phi_2,\Phi_3,\Phi_4 $ & \cr
\tablespace
\tablespace
}
\hrule}\hskip 1pt \vrule}
\vskip-98pt\hskip120pt
\vbox{$$\eqalign{\Phi_1 =&\, (0,0)\cr
            \Phi_2 =& \left({1\over 5},-{1\over 25}\right)\cr
            \Phi_3 =& \left({1\over 5},{1\over 25}\right)\cr
            \Phi_4 =& \left(-{1\over 10},0\right)\cr
}$$}
\hbox{\ Table 7: Fusion at $ c=-{6 \over 5} $}
}
\msk
\overfullrule3pt
How can this diagram be explained in our approach? First of all we cannot
reproduce the fusion coefficients themselves, since all we demand is
that the respective three point function does not vanish, i.e.~we
find necessary conditions for the appearance of a particular field in the
fusion, see also eq.(5.14).
To make the calculations explicit, we compute all data that we
need in order to apply eq.(5.18) to the case of $c=-6/5$.
The equations under consideration are
$$\eqalign{  \beta^2{\hat l_{ci}}  +
            {\hat m_{ci}}  =&\,
             \beta^2\left(\hat l_{di}+\hat l_{ei}\right) +
            {\hat m_{di}+\hat m_{ei}} \cr  &+
                 n_i\,\beta^2 + \hat n_i\ ,\cr} $$
as well as
$$\eqalign{  \beta^2{\hat l_{ci}}  +
            {\hat m_{ci}}  =&\,
             \beta^2\left( \hat l_{di}-\hat l_{ei}\right) +
            \left({\hat m_{di}-\hat m_{ei}}\right) \cr  &-
                 n_i\,\beta^2 - \hat n_i \cr} $$
and
$$\eqalign{  \beta^2{\hat l_{ci}}  +
            {\hat m_{ci}}  =&\,
             \beta^2\left( \hat l_{ei}-\hat l_{di}\right) +
            \left({\hat m_{ei}-\hat m_{di}}\right) \cr  &-
                 n_i\,\beta^2 - \hat n_i\ .\cr} $$
It is important to realize that $\hat n_2$ is always zero and
$n_1, \hat n_1, n_2$ are non-negative. In table 8 we give
all necessary data:
\msk
\vbox{
\hbox{
\vrule \hskip 1pt
\vbox{ \offinterlineskip
\def\tablespace{height2pt&\omit&&\omit&&\omit&&\omit&&\omit&\cr}
\def\tablerule{ \tablespace
                \noalign{\hrule}
                \tablespace        }
\def\tablesmallspace{height1pt&\omit&&\omit&&\omit&&\omit&&\omit&\cr}
\def\tableinterrule{ \tablespace
                \noalign{\hrule}
		\tablesmallspace
		\noalign{\hrule}
                \tablespace        }
\hrule
\halign{&\vrule#&
\strut\hskip 4pt\hfil#\hfil\hskip 4pt\cr
\tablespace
& $\left(\Delta_T,\, \Delta_W^2\right)$  &&
$\left(0,\,0\right)$ &&
$\left({1\over 5},-{1\over 25}\right)$ &&
$ \left({1\over 5}, {1\over 25}\right)$ &&
$\left(-{1\over 10}, 0 \right)$
&\cr
\tablespace
\tableinterrule
& $\left(l_1,m_1,l_2,m_2 \right)$ &&
$(1,1,1,1)$ &&
$(2,2,1,1)$ &&
$(1,1,1,3)$ &&
$(1,3,1,1)$
& \cr
\tablespace
\tablerule
& $\left(k_1, k_2\right)$ &&
$(0,0)$ &&
$(0,-2) $ &&
$(1,1) $ &&
$(0,1) $ &
\cr
\tablespace
\tablerule
& $\left(\bar l_1,\bar m_1,\bar l_2,\bar m_2   \right)$ &&
(-1,-1,1,-1) &&
(-2,-2,1,-1) &&
(-1,-1,1,-3) &&
(-1,-3,1,-1) & \cr
\tablespace
\tablerule
& $\left( \bar k_1, \bar k_2 \right)$ &&
$(-1,3) $ &&
$(-1,5) $ &&
$(-2,2) $ &&
$(-1,2) $ & \cr
\tablespace
\tablespace
}
\hrule}\hskip 1pt \vrule}
\hbox{\ Table 8: Representations of fields at $ c=-{6 \over 5} $}
}
\msk
Here $\beta^2 = -5$ and for example the fusion $\Phi_1 \times \Phi_i
= \Phi_i $ is understood if one takes $n_1 = n_2 = \hat n_2 = 0$
and the unbared representation of the identity for the first two
equations, and the bared one for the last equation. The representation
for the field $\Phi_i$ is arbitrary, except that for the last
equation the representation of the RHS is the bared of the one of the LHS
(recall that the conjugation is involutive). The pattern of the choice of
representatives and the numbers of screening charges involved for
the first equation can be determined from the following table with the
understanding that the given values are sometimes but by no means always
unique:
\sk
\vbox{
\hbox{
\vrule \hskip 1pt
\vbox{ \offinterlineskip
\def\tablespace{height2pt&\omit&&\omit&&\omit&&\omit&&\omit&\cr}
\def\tablerule{ \tablespace
                \noalign{\hrule}
                \tablespace        }
\def\tablesmallspace{height1pt&\omit&&\omit&&\omit&&\omit&&\omit&\cr}
\def\tableinterrule{ \tablespace
                \noalign{\hrule}
		\tablesmallspace
		\noalign{\hrule}
                \tablespace        }
\hrule
\halign{&\vrule#&
\strut\hskip 4pt\hfil#\hfil\hskip 4pt\cr
\tablespace
& $\left(\Delta_T,\, \Delta_W^2\right)$  &&
$\left(0,\,0\right)$ &&
$\left({1\over 5},-{1\over 25}\right)$ &&
$\left({1\over 5}, {1\over 25}\right)$ &&
$\left(-{1\over 10}, 0 \right)$
&\cr
\tablespace
\tableinterrule
& $\left(0,\,0\right)$ &&
$(u,u,u,0,0,0) $ &&
$(u,u,u,0,0,0) $ &&
$(u,u,u,0,0,0) $ &&
$(u,u,u,0,0,0) $
& \cr
\tablespace
\tablerule
& $\left({1\over 5},-{1\over 25}\right)$ &&
$(u,u,u,0,0,0)  $ &&
${(u,b,b,0,0,0)\atop (u,b,u,0,1,1)} $  &&
$(b,a,a,0,0,1) $  &&
${(b,u,u,0,2,1)\atop (b,u,u,0,1,1)} $ &
\cr
\tablespace
\tablerule
&$ \left({1\over 5}, {1\over 25}\right)$ &&
$(u,u,u,0,0,0)  $ &&
$(b,a,a,0,0,1) $ &&
${(u,b,b,0,0,0) \atop (u,u,b,1,1,0)}  $ &&
${(b,u,u,0,2,1)\atop (u,u,b,1,3,0)} $ & \cr
\tablespace
\tablerule
& $\left(-{1\over 10}, 0 \right)$  &&
$(u,u,u,0,0,0)  $ &&
${(b,u,u,0,2,1)\atop (b,u,u,0,1,1)} $ &&
${(b,u,u,0,2,1)\atop (u,u,b,1,3,0)} $ &&
${{(u,b,b,0,0,0) \atop (u,b,u,0,1,1)}\atop
{(u,u,b,1,3,0)\atop (u,b,u,1,4,0)}}$& \cr
\tablespace
\tablespace
}
\hrule}\hskip 1pt \vrule}
\hbox{\ Table 9: Fusion pattern at $ c=-{6 \over 5} $}
}
\sk
Here an entry like $(b,u,u,0,2,1)$ means that the conjugated (bared)
charge $\epsilon$
is fused with the unbared $\delta$ and results into an unbared $\gamma$.
The last three entries denote the respective number of screening charges
$(n_1,\hat n_1,n_2)$. Multiple entries are ordered with respect to their
appearance in table 7.
\sk
As a result all non-zero fusion coefficients of table 7
can be reproduced. On the other hand fusion coefficients equal to zero are not
always predicted, which is a common failure of the
free field approach. One may ask if there exists trivially a solution
for $(n_1,\hat n_1,n_2)$, but as a counterexample it may serve
that e.g.~the fusion of $\Phi_2$ with itself forbids the appearance of
$\Phi_3$ and $\Phi_4$, as can easily be checked.
The whole argument involving the three point functions relies on the fact
that the two-point function has
$\langle 0|V_{\bar\gamma}\, V_\gamma|0\rangle$ as its only non-zero
representation. So in
$\langle 0|V_{\bar\gamma}\, V_\delta\, V_\epsilon\, Q_{\tilde\eta_i}
\dots Q_{\tilde\eta_j}|0\rangle $ the fusion of $V_\delta$ and
$V_\epsilon$ together with the screening charges is supposed to yield
$V_{\gamma}$. On the other hand in
$\langle 0|V_{\gamma}\, V_{\bar\delta}\, V_\epsilon\, Q_{\tilde\eta_i}
\dots Q_{\tilde\eta_j}|0\rangle$ the fusion of $V_{\bar\delta}$ and
$V_\epsilon$ yields $V_{\bar\gamma}$ and analogously for the third three
point function. In this way we understand the pattern of solutions
for the second and third equation given above. The number of screening
charges stays the same as in table 7.
A new aspect of all this is that starting from eq.(5.18) one cannot split
it into two equations by collecting terms with equal powers of $\beta$, which
can be done for the $A_n$-algebras for example.
\sk
Next we inspect the other known fusion a little closer [32].
%
\msk
\vbox{
\hbox{
\vrule \hskip 1pt
\vbox{ \offinterlineskip
\def\tablespace{height2pt&\omit&&\omit&&\omit&&\omit&&\omit&&\omit&&\omit&\cr}
\def\tablerule{ \tablespace
                \noalign{\hrule}
                \tablespace        }
\def\tablesmallspace{
height1pt&\omit&&\omit&&\omit&&\omit&&\omit&&\omit&&\omit&\cr}
\def\tableinterrule{ \tablespace
                \noalign{\hrule}
		\tablesmallspace
		\noalign{\hrule}
                \tablespace        }
\hrule
\halign{&\vrule#&
\strut\hskip 4pt\hfil#\hfil\hskip 4pt\cr
\tablespace
& $\left(\Delta_T,\, \Delta_W\right)$ &&
$\left(0,\,0\right)$ &&
$\left(0,{3\over 22}\right)$ &&
$\left({1\over 12}, -{1\over 24}\right)$ &&
$\left(-{1\over 6},{1\over 66}\right)$ &&
$\left({3\over 4}, {39\over 88}\right)$ &&
$\left(-{1\over 4}, -{1\over 88}\right)$
&\cr
\tablespace
\tableinterrule
& $\left(0,\,0\right)$ &&
$\Phi_1$ &&
$\Phi_2$ &&
$\Phi_3$ &&
$\Phi_4$ &&
$\Phi_5$ &&
$\Phi_6$
& \cr
\tablespace
\tablerule
& $\left(0,{3\over 22}\right)$ &&
$ \Phi_2 $ &&
$ \Phi_1, \Phi_2, \Phi_4 $ &&
$ \Phi_3, \Phi_6 $ &&
$ \Phi_2, \Phi_4 $ &&
$ \Phi_6 $ &&
$ \Phi_3,  \Phi_5, \Phi_6 $ &
\cr
\tablespace
\tablerule
& $\left({1\over 12}, -{1\over 24}\right)$  &&
$ \Phi_3 $ &&
$ \Phi_3\Phi_6 $ &&
$ \Phi_1,\Phi_2,\Phi_4 $ &&
$ \Phi_3,\Phi_5,\Phi_6 $ &&
$ \Phi_4 $ &&
$ \Phi_2,\Phi_4 $ &
\cr
\tablespace
\tablerule
& $\left(-{1\over 6},{1\over 66}\right)$ &&
$ \Phi_4 $ &&
$ \Phi_2 \Phi_4$ &&
$ \Phi_3,\Phi_5,\Phi_6 $ &&
$ \Phi_1,\Phi_2,\Phi_4 $ &&
$ \Phi_3 $ &&
$ \Phi_3,\Phi_6 $ &
\cr
\tablespace
\tablerule
& $\left({3\over 4}, {39\over 88}\right)$ &&
$ \Phi_5 $ &&
$ \Phi_6 $ &&
$ \Phi_4 $ &&
$ \Phi_3 $ &&
$ \Phi_1 $ &&
$ \Phi_2 $ &
\cr
\tablespace
\tablerule
& $\left(-{1\over 4}, -{1\over 88}\right)$  &&
$ \Phi_6 $ &&
$ \Phi_3, \Phi_5, \Phi_6, $ &&
$ \Phi_2,\Phi_4 $ &&
$ \Phi_3,\Phi_6 $ &&
$ \Phi_2 $ &&
$ \Phi_1,\Phi_2,\Phi_4 $ &
\cr
\tablespace\tablespace
}
\hrule}\hskip 1pt \vrule}
\hbox{\ Table 10: Fusion at $ c=-{12 \over 7} $}
}
$$
\eqalign{\Phi_1 =& (0,0), \cr
         \Phi_2 =& (0,{3\over 22}), \cr}
\quad
\eqalign{\Phi_3 =& \left({1\over 12}, -{1\over 24}\right), \cr
         \Phi_4 =& \left(-{1\over 6},{1\over 66}\right), \cr}
\quad
\eqalign{\Phi_5 =& \left({3\over 4}, {39\over 88}\right),\cr
         \Phi_6 =& \left(-{1\over 4}, -{1\over 88}\right), \cr}
$$
This fusion can also be checked with our method, and we find agreement.
An interesting aspect of this model is the $Z_2$-symmetry which is respected
by the fusion, namely the fields group into even $\{\Phi_1,\Phi_2,\Phi_4\}$ and
odd $\{\Phi_3,\Phi_5,\Phi_6\}$ multiplets. This $Z_2$-symmetry is
most probably common to all models with even $q$. For reasons of brevity we
do not reproduce the analogues of the tables 8 and 9 here.
\sk
\subchhead{SW(3/2,~3/2,~2)}
\bsk
In order to repeat the previous analysis for the case of $D(2|1;\alpha)$
one has to look for all possible solutions of eqs.~(5.6). We did not
find $any$ solutions for the screening charges other than
$$ \tilde\eta_i = \beta\alpha_i \ ,\eq $$
even when we looked for ones not generic in the parameter $\alpha$.
This is interesting if one takes into account that $D(2|1;\alpha)$
is isomorphic to $Osp(4|2)$ for $\alpha=1$ or $-2$. The immediate
consequence of this fact is that there are no linearly dependent screening
charges, meaning that eq.(5.20) is trivial. As a result, one can not build
minimal models.
\sk
In order to gain more insight into this phenomenon we would like
to mention two further attempts to construction rational conformal
field theories for the $SW$ algebra algebra based on $D(2|1,\alpha)$
[33].
\sk
It is well known that rational conformal field theories exist only
for rational values of $c$. Since $c$ for the $SW$-algebra based on
$D(2|1,\alpha)$ is a continuous functions of $\alpha$ (see (4.12))
this implies a quantization of $\alpha$ for rational models. One
obvious trial for such a quantization is the choice $C_8 =0$
in eq.(4.21). Because of the presence of the additional parameter
$\beta$ the central charge is not yet fixed by this choice.
With this choice the two solutions for $SW(3/2,3/2,2)$ found in
[13] coincide. Therefore one would expect to be able to
write this $SW$-algebra as the direct sum of the $SW(3/2,2)$ discussed
at the beginning of section 4 and a super Virasoro algebra because
this is possible for the second solution to $SW(3/2,3/2,2)$ at
generic $c$ [13]. Vanishing self coupling constant $C_8$ implies
that the central charge of this $SW(3/2,2)$ is equal to $-{6 \over 5}$
which leads to a minimal model. This gives rise to the hope that
with an appropriate choice of the central charge of the additional
Super Virasoro algebra involved one can construct minimal models
of the $SW$ algebra based on $D(2|1,\alpha)$ in this manner.
Unfortunately, the change of basis from $SW(3/2,2) + SW(3/2)$
to a $SW(3/2,3/2,2)$ becomes singular precisely for vanishing
self coupling constant (compare also [27]). Therefore, the construction
explained above does not even give rise to a consistent $W$ algebra
and definitely not to rational conformal field theories.
\sk
In all known examples rational conformal field theories are
encoded in null fields appearing for particular values of the
central charge. This was also attempted along the lines of
[29] for the $SW$-algebra under consideration using a
non-covariant formulation. Looking for null fields at scale
dimension~4 (the lowest candidate dimension) one finds
null fields if one of the following relations is satisfied:
$$C_8^2 = - {(10 c -27)^2 c \over 486 (4 c + 21)}, \qquad
C_8^2 = {2 (10 c - 17)^2 c \over 27 (10 c - 7)}. \eq $$
In the first case one finds two null fields, in the
second case even three (at scale dimension 4).
Let $u$ be the eigenvalue of the upper component of the
dimension ${3 \over 2}$ generator in the representation,
$v$ the eigenvalue of the lower component of the
dimension 2 generator. Then all 8 conditions that were
evaluated are satisfied for
$$8 c^2 u - 10 c^2 v -180 c u + 27 c v - 27 c + 567 h
- 2268 u = 0 \eq $$
if one considers the case $C_8^2 = - {(10 c -27)^2 c \over 486 (4 c + 21)}$.
In particular it is not even possible to fix $c$ and
definitely not $h$. We should mention that because of
technical difficulties one possible type of conditions
was neglected. On the one hand, odds are low that this
would yield enough supplementary conditions. On the other
hand, all conditions factorize nicely with the same linear
factor which so far happened precisely if the model was
not rational but only degenerate. Note furthermore that from
(5.36) we indeed expect at most one relation in a degenerate
model.
\sk
In the case $C_8^2 = {2 (10 c - 17)^2 c \over 27 (10 c - 7)}$
the corresponding conditions are so complicated that it is
difficult to derive insight from them and therefore we
omit any details. Note that also in this case it is not
possible to fix $c$ as it would be necessary for a
rational model.
\sk
Finally we would like to notice that with the choice $C_8 = 0$
the determinant of all fields at scale dimension 4 is proportional
to some power of $c$ which is in accordance with the above remarks.
\sk
In summary, all three arguments we would have expected to
reveal rational models for our $SW(3/2,3/2,2)$ have failed to
do so. This is a strong hint that the $SW$-algebra based
on $D(2|1,\alpha)$ does indeed not have any rational models.
In part of the argumentation the parameter $\alpha$ played
an important role. Therefore it is not completely clear
if the reason for the absence of minimal models is this
parameter $\alpha$ or the purely fermionic root system.
Our general considerations at the beginning of this section
indicate, however, that the underlying reason most probably
is the purely fermionic root system, i.e.~the vanishing of
all diagonal elements of the Cartan matrix.
\sk
May be it is noteworthy that
exactly these minimal models were the original motivation of this work,
since it was an interesting perspective to study the appearance of the free
SSLA parameter $\alpha$ in the formulas.
\bsk
\chhead{Conclusions}
\bsk
On behalf of the preceding considerations we argue that the existence of
minimal models can be established from free field realizations
only if the Cartan matrix of the underlying SSLA has non-zero entries
on its diagonal. From ref.~[19]
we see that this is only the case for $Osp(2n+1|2n)$ and $Osp(2n-1|2n)$.
In general, series of minimal models cannot be built, and the cause of this can
be traced back to eq.(5.11). It may serve as a starting point, since in case
$\alpha_i^2\not =0$ the $l_i$ will be meaningless,
i.e.~we lose one parameter to describe degenerate theories. In the literature
we find the term ``not completely degenerate'', if there are not sufficiently
many null fields present to obtain differential equations which then determine
all $n$-point functions.
Furthermore, we observe from the explicit examples of $Osp(3|2)$ and
$D(2|1;\alpha)$ that for $\alpha_i^2 = 0$ we do not find a screening charge of
the form ${1\over\beta}\alpha_i$, while $\beta\alpha_i$ is always a solution.
These two screening charges (if they both exist) eventually give rise to the
same equation (5.12) (with the values of $l_i$ and $m_i$ exchanged), so that
for $\alpha_i^2 = 0$ this kind of ``pairing'' does not take place.
In a nutshell $\alpha_i^2=0$ implies no pairing and no minimal model
``in $\alpha_i$-direction''. From $Osp(3|2)$ we see that we need at least one
$\alpha_i^2\not =0$ in order to build minimal models, but then the other
charges are still involved (otherwise it would reduce to a Super-Virasoro
minimal model). It is interesting to note in this context that the
vanishing of all diagonal elements of $K_{ij}$ leads to an energy momentum
tensor which does not differ in its form from the classical one, see eq.(3.3).
So one may speculate that the system is in some way ``too classical'' to
generate minimal models, though the expressions for the conserved currents
differ from the corresponding classical ones.
Another approach to understand the absence of minimal models could be,
that supersymmetry is too restrictive to admit minimal models.
Because if one allows at the level of Toda Theories for bosonic
simple roots (which spoil supersymmetry) they have in general
non-zero entries on the corresponding Cartan matrices.
Most probably one can still generate $W$-algebras
(corresponding to an integrable field theory), so that these may then admit
minimal models.
\sk
The above discussion leads us to the conclusion that for the minimal models
of $SW(3/2,2)$ the conformal dimensions are almost known, in the sense
that they can be parameterized by eq.(5.29), but the selection rules for
the allowed parameters $l_1$, $m_1$ and $m_2$ have not been clarified
completely, yet. In particular the case of negative $p$ is not really
understood. A closely related question is why $p$ is of the form
$p = (-1)^q$, and whether it is correct to attribute the negative
sign to $p$, though it is favored to by the restriction of $l_1$ to one
and zero in that case.
\sk
It is important to realize, that all the results can be easily generalized
to the $SW$-algebras belonging to $Osp(2n+1|2n)$. The central charge
for minimal models is $c = {3\over 2}r + {12\over \beta^2} \rho_q^2$.
The conformal dimensions of multiplets are given by
$\Delta_T = -{1\over 2}\beta^2\eta^2 - \eta\cdot\rho_q $. Here
$\eta$ is of course again determined by eq.(5.13). The same thing applies
to the fusion rules.
\bsk
\subchhead{Acknowledgements}
\bsk
I would like to thank W.~Nahm, R.~Blumenhagen, M.~Flohr,
J.~Kellendonk, N.~Mohammedi, A.~Recknagel, M.~R\"osgen, M.~Terhoven and
R.~Varnhagen, for encouragement and fruitful discussions.
I am especially grateful to W.~Eholzer, A.~Honecker and R.~H\"ubel
for reading the manuscript, providing the examples of section five and
further useful discussions. Also I would like to thank the computer centre
of the Max-Planck-Institut f\"ur Mathematik Bonn
for access to to their Unix-cluster
and the Rechenzentrum der Universit\"at T\"ubingen for access to their convex.
Finally, I would also like to thank the Max-Planck-Institut f\"ur Physik
komplexer Systeme Dresden and Prof.~Fulde for support during the final stages
of this work.
\vfill
\eject

\subchhead{Appendix A: Rules for the computation of covariant OPE's and NOP's}
\sk
Basic for the implementation of the package "OPESUSY" which computes
$N$=1 supersymmetric OPE's and reduces normal ordered products to
a standard form are the following rules.
The notation is set by
$$ A(Z_1)\circ B(Z_2) = \sum_{n=0}^{max} {[AB]_n(Z_2) \over
{Z_{12}}^{n/2}}\ , \eqno{(A.1)}  $$
where the power of $Z_{12}$ was defined in eq.(2.3).
The definition of the integral over Grassmann numbers is
$$ \int d\theta\, 1 = 0,\qquad \int d\theta\, \theta = 1\ . \eqno{(A.2)} $$
So the supersymmetric generalization of the Cauchy theorem is
$${1\over 2 \pi i} \oint_{C_{z_2}}dz_1d\theta_1{\Phi(Z_1)\over {Z_{12}}^{n/2}}
={1\over 2 \pi i} \oint_{C_{z_2}}dz_1d\theta_1 {Z_{12}}^{-n/2}\Phi(Z_1)
={1 \over [{n-1\over 2}]!} D^{n-1} \Phi(Z_2)\ .\eqno{(A.3)}$$
Note the hereby defined order for a fraction, in case $\Phi$ and
${Z_{12}}^{n/2}$ are both fermionic.
Subsequently a complete set of rules is given, meaning that it is sufficient
to compute arbitrary OPE's and reduce normal ordered products to
standard order, provided some ``basic'' OPE's are given
$$ [BA]_q = (-1)^{AB} \sum_{{n=0\atop n\, even, if\, q\, even}}^{max}
{(-1)^{[{n+q\over 2}]} \over [{n \over 2}]!}\ D^n[AB]_n
\eqno{(A.4)}$$
The important special case of two identical fermionic fields is given by:
$$
[AA]_0 = -{1\over 2} \sum_{n=1}^{max} {(-1)^n \over n!}
\ D^{2n}[AA]_{2n}  \eqno{(A.5)}   $$  \sk
The analogue to Wick's theorem is:
$$ \eqalign{[A[BC]_0]_q &= (-1)^{AB} (-1)^{(2\,q \Delta_B)}\ [B[AC]_q]_0 \ +
\ [[AB]_qC]_0  \cr
&+ \sum_{{n=0\atop q-n\ odd}}^{q-2}
{[{q-1\over 2}]!\over[{q-n-1\over 2}]! [{n \over 2}]!}
\ [[AB]_{n+1}C]_{q-n-1} \cr
[A[BC]_0]_0 &= (-1)^{AB} [B[AC]_0]_0
 - \sum_{n=1}^{max} {(-1)^n  \over n!}\ [D^{2n} [AB]_{2n}C]_0
\phantom{()} \cr
[A[AC]_0]_0 &=-{1 \over 2} \sum_{n=1}^{max} {(-1)^n \over [{n \over 2}]!}
\ [D^{2n} [AA]_{2n}C]_0
\hbox{\ , A fermionic.}  \cr }\eqno{(A.6)}
$$  \bsk
\vfill\eject
\currentch{Appendix B: Solutions for the spin-2 field}
\subchhead{B. Solutions for the spin-2 field}
\bsk
Although the solution given below belongs to a linear system
of equations, it took a few hours of spark2 cpu-time to obtain it.
It reads
\overfullrule=1pt
$$\eqalign{
  U_1 =~& (V_{23}  (\beta^2 + \beta^2 \alpha) + V_{31}
 (4 \alpha - \beta^2 \alpha^2) +
      V_{32}  (-\beta^2 - 2 \beta^2 \alpha - \beta^2 \alpha^2))
                  /(4 \beta \alpha (1 + \alpha)),\cr
  U_{11} =~& (V_{23}  (-1 - \alpha) + V_{31}  \alpha^2 + V_{32}
  (1 + 2 \alpha + \alpha^2))
             /(4 \alpha (1 + \alpha)),\cr
  U_{112} =~& (\beta V_{31}  \alpha^2 + V_{23}  (-\beta - \beta \alpha) +
 V_{32}(\beta + 2 \beta \alpha + \beta \alpha^2))/(2 \alpha (1 + \alpha)),\cr
  U_{113} =~&(-(\beta V_{31}  \alpha^2) + V_{23}  (\beta + \beta \alpha) +
V_{32} (-\beta - 2 \beta \alpha - \beta \alpha^2))/(2 \alpha),\cr
  U_{12} =~&(V_{23}(1-\beta^2+\alpha -\beta^2\alpha)+V_{31}
 (-2\alpha -\alpha^2+\beta^2\alpha^2) + \cr
{}~&V_{32}
(1+\beta^2+2\beta^2\alpha-\alpha^2+\beta^2\alpha^2))/(\alpha(1 +\alpha)),\cr
  U_{123} =~& (V_{23}(-\beta -\beta\alpha)+V_{31}
(-2\beta\alpha -\beta\alpha^2)+V_{32}
        (-\beta - 2 \beta \alpha - \beta \alpha^2))/(2 (1 + \alpha)),\cr
U_{13} =~& (V_{23}(-1 - 3 \alpha - 2 \alpha^2 + \beta^2(1  + 2 \alpha
+ \alpha^2)) +\cr
{}~&V_{32} (-1 -2\alpha -\alpha^2 -\beta^2(1 -3 \alpha
-3 \alpha^2- \alpha^3))\cr &+
V_{31}  (2 \alpha + \alpha^2 - \beta^2 \alpha^2 - \beta^2 \alpha^3))/(\alpha
  (1 + \alpha)),\cr
U_2 =~&(-V_{31}(\beta\alpha)^2 + V_{23}\beta^2(1+\alpha)+V_{32}
  (-4-\beta^2-4\alpha -2\beta^2 \alpha - \beta^2 \alpha^2))/
  (4 \beta(\alpha + \alpha^2)),\cr
  U_{212}=~&(-(\beta V_{31}\alpha^2)+V_{23}(\beta +\beta\alpha)+
V_{32}(-\beta -2\beta\alpha
            -\beta\alpha^2))/(2 \alpha (1 + \alpha)),\cr
  U_{213}=~&(\beta V_{31}\alpha^2+V_{23}(\beta +\beta\alpha)+V_{32}
(-\beta +\beta\alpha^2))/(2\alpha),\cr
  U_{22}=~&(V_{23}(-1-\alpha)+V_{31}\alpha^2+V_{32}(1+2\alpha +\alpha^2))/
(4\alpha(1+\alpha)),\cr
  U_{223} =~& (\beta V_{31}  \alpha^2 + V_{23}  (-\beta - \beta \alpha) +
V_{32} (\beta + 2 \beta \alpha + \beta \alpha^2))/(2 (1 + \alpha)),\cr
  U_{23} =~&(V_{23}(1+3\alpha-\beta^2\alpha +2\alpha^2-\beta^2\alpha^2)+
V_{31}(-\alpha^2+\beta^2\alpha^3)
       \cr&+ V_{32}(-1+\beta^2\alpha +\alpha^2+2\beta^2\alpha^2+
\beta^2\alpha^3))/(\alpha(1+\alpha)),\cr
U_3 =~& (-\beta^2 V_{31}  \alpha^2 + V_{23}
(-4 + \beta^2 - 4 \alpha + \beta^2 \alpha) +
    V_{32} \beta^2 (-1 - 2 \alpha - \alpha^2))/(4\beta\alpha(1 + \alpha)),\cr
  U_{312} =~& (-(\beta V_{31}  \alpha^2) + V_{23}
(-\beta - 3 \beta \alpha - 2 \beta \alpha^2) +
      V_{32}  (\beta + 2 \beta \alpha + \beta \alpha^2))/
(2 \alpha (1 + \alpha)),\cr
  U_{313} =~&(\beta V_{31}\alpha^2+V_{23}(-\beta -\beta\alpha)+
V_{32}(\beta +2\beta\alpha +\beta\alpha^2))
         /(2 \alpha),\cr
  U_{323} =~&(-(\beta V_{31}\alpha^2)+V_{23}(\beta +\beta\alpha)+
V_{32}(-\beta -2\beta\alpha -
             \beta\alpha^2))/(2 (1 + \alpha)),\cr
 U_{33} =~&(V_{23}(-1-\alpha)+V_{31}\alpha^2+V_{32}(1+2\alpha+\alpha^2))/
(4\alpha(1+\alpha)),\
 V_{11} =~ 0, \
 V_{12} =~ V_{32} /\alpha, \cr
 V_{13} =~& -V_{23}  (\beta + \beta \alpha)/(\beta \alpha),\quad
 V_{21} =~ -\beta\, V_{31} /(\beta + \beta \alpha), \quad
 V_{22} =~ 0, \qquad V_{33} =~ 0.\cr}\eqno{(B.1)}
$$
\overfullrule=4pt
The solutions contain three free parameters corresponding to an
admixture of the covariant derivatives of $T$ and $M$ and the yet
undetermined normalization of a the new spin-2 field $U$.
The condition of primarity again fixes these three free parameters up to one
$$\eqalign{V_{31} =~& -(1 +\alpha)(2 +\beta^2 \alpha)\,V_{23}
/(2\alpha^2 + \beta^2\alpha^2)\ , \cr
  V_{32} =~& (-2 + \beta^2 +\beta^2 \alpha)\,V_{23}/
((2 + \beta^2)(1 + \alpha))\cr\ .}\eqno{(B.2)}
$$
\vfill\eject
\currentch{References}
\baselineskip9pt
\vsize9.0in
\vskip-13pt
\par\refrm
\ref [1]!A.A. Belavin,A.M. Polyakov and A.B. Zamolodchikov:!
  Infinite Conformal Symmetry in Two-dimen\-sional Quantum Field Theory!
  Nucl.Phys.!B241!(1984) 333
\ref [2]!V.S. Dotsenko, V.A. Fateev:!
  Conformal Algebra and Multipoint Correlation Functions in 2D
  Statistical Models!
  Nucl.Phys.!B240!(1984) 312
\ref [3]!G. Felder:!
  BRST Approach to Minimal Models! Nucl.Phys.!B317!(1989) 548
\ref [4]!L. Alvarez-Gaumez and Ph. Zaugg:!
  Structure Constants in the N=1 Superoperator Algebra!
  Ann. Phys. !215!(1992) 171
\ref [5]!M.A. Bershadsky, V.G. Knizhnik and M.G. Teitelman:!
  Superconformal Symmetry in two Dimensions!
  Phys.Lett.!B151!(1985)
\ref [6]!K. Sugiyama:!
  BRST Analysis of N=2 Superconformal Minimal Unitary Models
      in Coulomb Gas Formalism! Preprint!{}!{UT-646}
\ref [7]!P. Bouwknegt, K. Schoutens:! W-Symmetry in Conformal Field Theory!
         Phys.Rep.!223!(1993)
\ref [8]!V.A. Fateev and S.L. Lykyanov:!
   The Models of Two-dimensional Conformal Quantum Field Theory with
   $\hbox{\refrm Z}_n$ Symmetry!
  Int.J.Mod.Phys.!A3!(1988) 507
\ref [9]!V.~Kac, M.~Wakimoto, E.~Frenkel:! {Characters and Fusion Rules
for W-algebras via Quantized Drinfeld-Sokolov-Reduction}! CMP!{147}!(1992) 295
\ref [10]!S. Penati and D. Zanon:!
  Quantum Symmetries in Supersymmetric Toda Theories!
  Rome String Workshop!{}!(1992) 450
\ref [11]!K. Thielemans:!
  A Mathematica${}^{TM}$-Package for Computing Operator Product
        Expansions!
  Int.J. Mod.Phys.!C!Vol.2, No.3 (1991) 787
\ref [12]!R. Blumenhagen, W. Eholzer, A. Honecker and R. H\"ubel:!
  New N=1 Extended Superconformal Algebras with two and three Generators!
  Int.J.Mod.Phys.!A7!(1992) 7841
\ref [13]!R. Blumenhagen:!
  Covariant Construction of N=1 Super W-Algebras!
  Nucl.Phys.!B381!(1992) 641
\ref [14]!S. Schrans:!
        Extensions of Conformal Invariance in Two-dimensionsal
        Quantum Field Theory! Ph.D. thesis!{}! Leuven University 1991
\ref [15]!P. Mansfield and B. Spence:!
  Toda-Theories, W-Algebras, and Minimal Models!
  Nucl.Phys.!B362!(1991) 294
\ref [16]!P. Mansfield:!
  Conformally Extended Toda Theories! Phys.Lett.!B242!(1990) 387
\ref [17]!B. Feigin and E. Frenkel:!
  Free Field Resolutions in Affine Toda Theories!
  Phys.Lett.!B276!(1992) 79
\ref [18]!S. Komata, H. Nohara and K. Mohri:!
  Classical and Quantum Extended
  Superconformal Algebras!
   Nucl.Phys.!B359!(1991) 168
\ref [19]!J. Evans and T. Hollowood:!
  Supersymmetric Toda Field Theories!
  Nucl.Phys.!B352!(1991) 723
\ref [20]!H. Nohara and K. Mohri:!
  Extended Superconformal Algebras from
  Super Toda Field Theory!
   Nucl.Phys.!B349!(1991) 253
\ref [21]!H.C. Liao and P. Mansfield:!
  Light-cone Quantization of the Super-Liouville-Theory!
  Nucl.Phys.!B344!(1990) 696
\ref [22]!L. Frappat, E. Ragoucy and P. Sorba:!
  (Super) W-algebras from non Abelian
  (super) Toda theories! hep-th!\null!9209123
\ref [23]!L.A. Ferreira, J.F. Gomes, R.M. Ricotta and A.H. Zimmerman:!
  Supersymmetric Construction of W-Algebras from  Super Toda and WZNW
  Theories! Int.J.Mod.Phys.!A7! (1992) 7713
\ref [24]!H.G. Kausch and G.M.T. Watts:!
  Quantum Toda Theory and the Casimir Algebra of B2 and C2!
  Int.J.Mod.Phys.!A7!(1992) 4175
\ref [25]!H.G.\ Kausch and G.M.T.\ Watts:!
  Duality in Quantum Toda Theory and W-Algebras!
  Nucl.Phys.!B386!(1992) 166
\ref [26]!J.M. Figurea-O'Farrill, S. Schrans!
	The Conformal Bootstrap and Super W-Algebras!
	Int.J.Mod.Phys.! A7! (1992) 591
\ref [27]!W.\ Eholzer, M.\ Flohr, A.\ Honecker, R.\ H{\"u}bel, W.\ Nahm,
          R.\ Varnhagen!
   Representations of W-Algebras with two Generators
   and New Rational Models!
   Nucl.Phys.!\ B383! (1992) 249
\ref [28]!V.G. Kac:! Lie Superalgebras
          !Adv.Math.!26!(1977) 8
\ref [29]!W. Eholzer, A. Honecker and R. H\"ubel:!
  Representations of N=1 Extended Superconformal
  Algebras with two Generators! Mod.Phys.Lett.!{A8}!(1993) 725
\ref [30]!R.\ H\"ubel, private communication!{}!{}!{}!{}
\ref [31]!W. Eholzer and R. H\"ubel:!Fusion Algebras of Fermionic Rational
      Conformal Field Theories via a Generalized Verlinde Formula!
      Nucl.Phys.!B414!(1994) 348
\ref [32]!M.Flohr:! W-Algebras, New Rational Models and Completeness of the c=1
       Classification!
       Commun.Math.Phys.{\refrm 175}!{}!(1993) 179-212
\ref [33]!A.\ Honecker, private communication!{}!{}!{}!{}

\bye